\documentclass[10pt]{article}

\usepackage[utf8]{inputenc}
\usepackage{tikz}
\usetikzlibrary{shapes,arrows}
\usepackage[none]{hyphenat} % para no cortar las palabras
\usepackage{amsmath}
\usepackage{amssymb}
\usepackage{amsfonts}
\usepackage{multirow}
\usepackage{natbib}
\usepackage{enumerate}
\usepackage{hyperref}
\usepackage{graphicx}
\hypersetup{colorlinks=true, linkcolor=blue, urlcolor=blue, citecolor=red}
\usepackage{listings}
\usepackage{color}
\usepackage{epstopdf}
\usepackage{rotating}
\usepackage{algorithm}
\usepackage{algorithmic}

\usepackage[top=2cm, bottom=2cm, left=2cm, right=2cm]{geometry}

\newtheorem{proposition}{Proposition}
\newtheorem{corollary}{Corollary}
\newtheorem{proof}{Proof}
\title{\bf Generalized Ridge Regression: Biased Estimation for Multiple Linear Regression Models}
\author{
    \bf{Rom\'an Salmer\'on G\'omez}\thanks{Professor, Department of Quantitative methods for economics and business, University of Granada, Spain (e-mail: romansg@ugr.es).}
    \and
    \bf{Catalina Garc\'ia Garc\'ia}\thanks{Professor, Department of Quantitative methods for economics and business, University of Granada, Spain (e-mail: cbgarcia@ugr.es).}
    \and 
    \bf{Guillermo Hortal Reina}\thanks{PhD student. University of Granada. Spain  (e-mail: ghorrei@correo.ugr.es).}
}
\date{\today}

\begin{document}

\maketitle

\maketitle

\begin{abstract}
 When the regressors of a econometric linear model are nonorthogonal, it is well known that their estimation by ordinary least squares can present various problems that discourage the use of this model. The ridge regression is the most commonly used alternative; however, its generalized version has hardly been analyzed. The present work addresses the estimation of this generalized version, as well as the calculation of its mean squared error, goodness of fit and bootstrap inference. \\
 \textbf{Keywords:} generalized ridge regression, mean squared error, norm, goodness of fit. 
\end{abstract}

\section{Introduction}

    Given the following multiple linear regression model
    \begin{equation}
        \label{modelo}
        \mathbf{Y} = \mathbf{X}  \boldsymbol{\beta} + \mathbf{u},
    \end{equation}
     where $\mathbf{X}$ is an $n \times p$ matrix with full rank, $\boldsymbol{\beta}$ is a vector with unknown parameters (to be estimated), $E[\mathbf{u}] = \mathbf{0}$ and $E[\mathbf{u} \mathbf{u}^{t}] = \sigma^{2}  \mathbf{I}$ (being $\mathbf{I}$ is the identity matrix), the ridge estimation proposes adding a small positive quantity to the diagonal of the matrix $\mathbf{X}^{t} \mathbf{X}$ to mitigate the effects of nonorthogonality in the regression model, leading to biased estimators with a mean squared error lower than that obtained from the ordinary least squares (OLS). Although the first references to the ridge regression date to the 1960s (see \cite{Hoerl1962}, \cite{Hoerl1964} and \cite{HoerlKennard1968}), it was not until the works of \cite{HoerlKennard1970a,HoerlKennard1970b} that this technique was developed in depth. \cite{hoerl2020ridge} presented an interesting paper reviewing the origins of ridge regression, its developments and extensions. \cite{hastie2020ridge} have also collected some of the developments and applications of ridge regression within the field of applied statistics. Recently, \cite{zhang2022ridge} stated that ridge regression may be worth another look since it may offer some advantages over the Lasso (\cite{tibshirani1996regression}), for example it can be easily computed with a closed-form expression.

     Precisely the fact of having a closed-form expression has opened a line of research on how to theoretically justify the increase in the diagonal of matrix $\mathbf{X}^{t} \mathbf{X}$ has been a particular research line in the ridge regression literature. In this sense, \cite{piegorsch1989early} stated that \textit{finding a theoretically optimal basis for the ridge procedure has been a lengthy process (\cite{rolph1976choosing}, \cite{strawderman1978minimax}, \cite{casella1980minimax}), and it is still not fully developed}. Examining that theoretical justification, \cite{HoerlKennard1970a} indicated that the ridge estimator presented a contact point with other approximations in regression analysis and at least three of them should be commented:
    \begin{itemize}
        \item  The Stein estimator (\cite{stein1960multiple}).
        \item A Bayesian approach to regression (\cite{jeffreys1998theory} and \cite{raiffa1961applied}).
        \item Constrained maximization (\cite{balakrishnant1963operator}).
    \end{itemize}

    The ad hoc solution presented by \cite{HoerlKennard1970a} and \cite{HoerlKennard1970b} to the collinearity presented in the design matrix has been justified post hoc. We present a brief review of the justification provided by the scientific literature:
    \begin{itemize}
        \item Minimization of the ridge loss function,  \cite{harville1998matrix} and \cite{fletcher2013practical}, similar to penalized models. The most common penalty term is the bridge penalty term (\cite{frank1993statistical}, \cite{fu1998penalized}):
            \[P(\boldsymbol{\beta})=\sum_{j=1}^{p}|\beta_j|^\alpha,\]
            where $\alpha>0$ is an adjustment parameter. For $\alpha=2$, the ridge regression is obtained (\cite{HoerlKennard1970a} and \cite{HoerlKennard1970b}); while for $\alpha=1$, the Lasso estimator (\cite{tibshirani1996regression}) is obtained. Penalties with $\alpha<1$ have also been called soft thresholding (\cite{donoho1995adapting}; \cite{klinger1998hochdimensionale}). Recently, \cite{zou2005regularization} proposed elastic net regularization by using a penalty term combining the ridge and Lasso penalties:
            \[P(\boldsymbol{\beta})=\lambda_1\sum_{j=1}^{p}|\beta_j|+\lambda_2\sum_{j=1}^{p}|\beta_j|^2, \quad \lambda_1, \lambda_2>0.\]
            These methods are applied not only to treat multicollinearity but also for variable selection.
        \item The ridge regression also has a close connection to the Bayesian linear regression. This idea appeared in the works of \cite{HoerlKennard1970a} and \cite{Marquardt1970}. A posterior formalization is found in the paper of \cite{ljndley1972bayes}. %Numerous contributions exist since this connection was already provided \cite{HoerlKennard1970a} by using the references of \cite{jeffreys1998theory} and \cite{raiffa1961applied}.
    \end{itemize}

    In any case, \cite{HoerlKennard1970a} presented a general way to obtain the ridge estimation based on the decomposition of matrix $\mathbf{X}^{t} \mathbf{X}$ in its canonical form. Because $\mathbf{X}^{t} \mathbf{X}$ is a symmetric positive definite matrix, it is verified that there is an orthogonal matrix $\boldsymbol{\Gamma}$ (this is to say $\boldsymbol{\Gamma} \boldsymbol{\Gamma}^{t} = \mathbf{I} = \boldsymbol{\Gamma}^{t}  \boldsymbol{\Gamma}$) and a diagonal matrix $\boldsymbol{\Lambda}$ (both with $p \times p$ dimensions) such that $\mathbf{X}^{t} \mathbf{X} = \boldsymbol{\Gamma}  \boldsymbol{\Lambda}  \boldsymbol{\Gamma}^{t}$.
    Matrix $\boldsymbol{\Gamma}$ contains the eigenvectors of $\mathbf{X}^{t} \mathbf{X}$ and $\boldsymbol{\Lambda}$ the eigenvalues (which are real positives).

    Thus, given the model (\ref{modelo}), its canonical version is expressed as $\mathbf{Y} = \mathbf{Z}  \boldsymbol{\xi} + \mathbf{u}$, where $\mathbf{Z} = \mathbf{X}  \boldsymbol{\Gamma}$ and $\boldsymbol{\xi} = \boldsymbol{\Gamma}^{t}  \boldsymbol{\beta}$. In this case, the OLS estimator of $\boldsymbol{\xi}$ is $\widehat{\boldsymbol{\xi}} = \left( \mathbf{Z}^{t} \mathbf{Z} \right)^{-1} \mathbf{Z}^{t} \mathbf{Y} = \boldsymbol{\Lambda}^{-1} \mathbf{Z}^{t} \mathbf{Y}$. Then, the general ridge estimator is defined as:
    \begin{equation}
        \label{est_ridge_cano}
        \widehat{\boldsymbol{\xi}} (\mathbf{K}) = \left( \boldsymbol{\Lambda} + \mathbf{K} \right)^{-1} \mathbf{Z}^{t} \mathbf{Y},
    \end{equation}
    where $\mathbf{K} = diag(k_{1}, k_{2}, \dots, k_{p})$ being $k_{i} \geq 0$ for $i=1,\dots,p$. Note that following \cite{HoerlKennard1970a}, the optimal values for $k_{i}$ are $k_{i} = \sigma^{2}/\xi_{i}^{2}$, where $\xi_{i}$ are the elements of $\boldsymbol{\xi}$.

    Due to $\boldsymbol{\xi} = \boldsymbol{\Gamma}^{t}  \boldsymbol{\beta}$, the expression (\ref{est_ridge_cano}) can be expressed as:
    \begin{equation}
        \label{est_ridge}
        \widehat{\boldsymbol{\beta}}(\mathbf{K}) =  \boldsymbol{\Gamma}  \widehat{\boldsymbol{\xi}} (\mathbf{K}) = \boldsymbol{\Gamma} \left( \boldsymbol{\Gamma}^{t} \mathbf{X}^{t} \mathbf{X} \boldsymbol{\Gamma} + \mathbf{K} \right)^{-1} \boldsymbol{\Gamma}^{t} \mathbf{X}^{t} \mathbf{Y} = \left( \mathbf{X}^{t} \mathbf{X} + \boldsymbol{\Gamma}  \mathbf{K}  \boldsymbol{\Gamma}^{t} \right)^{-1} \mathbf{X}^{t} \mathbf{Y}.
    \end{equation}

   However, in the paper of \cite{HoerlKennard1970b} (page 70), the expression provided is:
    \begin{equation}
        \label{est_ridge_HK}
        \widehat{\boldsymbol{\beta}}(\mathbf{K}) = \left( \mathbf{X}^{t} \mathbf{X} + \mathbf{K} \right)^{-1} \mathbf{X}^{t} \mathbf{Y},
    \end{equation}
    which differs from the one obtained in expression (\ref{est_ridge}).

    It is true that in the particular case in which $k_{1} = k_{2} = \dots = k_{p} = k \geq 0$, i.e. $\mathbf{K} = k  \mathbf{I}$, expressions (\ref{est_ridge}) and (\ref{est_ridge_HK}) coincide because $\boldsymbol{\Gamma}$ is an orthogonal matrix and consequently $\boldsymbol{\Gamma}^{t}  \boldsymbol{\Gamma} = \mathbf{I}  = \boldsymbol{\Gamma}  \boldsymbol{\Gamma}^{t}$. Thus, in this case (which is universally used when the ridge regression is estimated), there is no contradiction. However, to analyze the generalized version of the ridge regression, the expression (\ref{est_ridge}) should be used instead of (\ref{est_ridge_HK}).

    The focus of this paper is to analyze this generalized version of the ridge regression: Section \ref{estimation} analyzes the properties of the estimator $\widehat{\boldsymbol{\beta}}(\mathbf{K})$ given in (\ref{est_ridge}), its norm (Section \ref{norm}), the mean squared error (Section \ref{mse}) and the goodness of fit (Section \ref{R2}) paying special attention to the particular case when $\mathbf{K} = diag(0, \dots, k_{l}, \dots, 0)$ with $k_{l} > 0$, $l=1,\dots,p$ since, as will be seen, it has advantages over the one usually used where $\mathbf{K} = k  \mathbf{I}$.
    Section \ref{matrix_mse} analyzes the performance of the proposed estimator under the root mean squared error matrix criterion while Section \ref{boots} proposes the implementation of inference using bootstrap methodology.
    Finally, Section \ref{example} illustrates the contribution of this paper with the example of \cite{GormanToman} used by \cite{HoerlKennard1970b}, and Section \ref{conclusion} summarizes the main conclusions of the work.

\section{Estimation properties}
    \label{estimation}

    This section analyzes the properties of the estimator $\widehat{\boldsymbol{\beta}}(\mathbf{K})$ given in (\ref{est_ridge}) and shows, among other questions, that it is biased. It is calculated as its matrix of variances and covariances and its trace. The augmented model that leads to this estimator is also analyzed.

    Thus, due to $\mathbf{X}^{t} \mathbf{X} = \boldsymbol{\Gamma}  \boldsymbol{\Lambda}  \boldsymbol{\Gamma}^{t}$, it is obtained that:
    \begin{equation}
        \label{beta_K}
        \widehat{\boldsymbol{\beta}}(\mathbf{K}) = \left( \boldsymbol{\Gamma}  \boldsymbol{\Lambda}  \boldsymbol{\Gamma}^{t} + \boldsymbol{\Gamma}  \mathbf{K}  \boldsymbol{\Gamma}^{t} \right)^{-1} \mathbf{X}^{t} \mathbf{Y} = \boldsymbol{\Gamma}  \boldsymbol{\Omega}  \boldsymbol{\Gamma}^{t}  \boldsymbol{\alpha} = \boldsymbol{\Gamma}  \boldsymbol{\Omega}  \boldsymbol{\delta},
    \end{equation}
    where:
    $$\boldsymbol{\alpha} = \mathbf{X}^{t} \mathbf{Y}, \quad
    \boldsymbol{\Omega} = \left( \boldsymbol{\Lambda} + \mathbf{K} \right)^{-1} = diag \left( \frac{1}{\lambda_{1}+k_{1}}, \dots, \frac{1}{\lambda_{p}+k_{p}} \right), \quad
    \boldsymbol{\delta} = \boldsymbol{\Gamma}^{t}  \boldsymbol{\alpha}.$$

    Definitely:
    \begin{equation}
        \label{est_ridge_i}
        \widehat{\boldsymbol{\beta}}(\mathbf{K}) = \boldsymbol{\Gamma}  \left(
        \begin{array}{c}
            \frac{\delta_{1}}{\lambda_{1}+k_{1}} \\
            \vdots \\
            \frac{\delta_{p}}{\lambda_{p}+k_{p}}
        \end{array} \right) \rightarrow \widehat{\boldsymbol{\beta}}(\mathbf{K})_{i} = \gamma_{i}  \left(
        \begin{array}{c}
            \frac{\delta_{1}}{\lambda_{1}+k_{1}} \\
            \vdots \\
            \frac{\delta_{p}}{\lambda_{p}+k_{p}}
        \end{array} \right) = \sum \limits_{j=1}^{p} \frac{\gamma_{ij}  \delta_{j}}{\lambda_{j}+k_{j}},
    \end{equation}
    where $\gamma_{i}$ is row $i$ of matrix $\boldsymbol{\Gamma}$ and $\widehat{\boldsymbol{\beta}}(\mathbf{K})_{i}$ is element $i$ of $\widehat{\boldsymbol{\beta}}(\mathbf{K})$, $i=1,\dots,p$.
    When $k_{j} \rightarrow +\infty$ for all $j$, it is verified that $\widehat{\boldsymbol{\beta}}(\mathbf{K})_{i} \rightarrow 0$.

    Furthermore, because the OLS estimator of model (\ref{modelo}) is $\widehat{\boldsymbol{\beta}} = \left( \mathbf{X}^{t} \mathbf{X} \right)^{-1} \mathbf{X}^{t} \mathbf{Y}$, $\widehat{\boldsymbol{\beta}}(\mathbf{K}) = \mathbf{W}_{K}  \widehat{\boldsymbol{\beta}}$ where $\mathbf{W}_{K} = \left( \mathbf{X}^{t} \mathbf{X} + \boldsymbol{\Gamma}  \mathbf{K}  \boldsymbol{\Gamma}^{t} \right)^{-1} \mathbf{X}^{t} \mathbf{X}$. In this case, $E \left[ \widehat{\boldsymbol{\beta}}(\mathbf{K}) \right] = \mathbf{W}_{K}  \boldsymbol{\beta} \not= \boldsymbol{\beta}$ unless $\mathbf{W}_{K} = \mathbf{I}$.
    In addition, due to $var \left( \widehat{\boldsymbol{\beta}} \right) = \sigma^{2}  \left( \mathbf{X}^{t} \mathbf{X} \right)^{-1}$, it is verified that:
    \begin{equation}
        var \left( \widehat{\boldsymbol{\beta}}(\mathbf{K}) \right) = \mathbf{W}_{K}  var \left( \widehat{\boldsymbol{\beta}} \right)  \mathbf{W}_{K}^{t} = \sigma^{2}  \left( \mathbf{X}^{t} \mathbf{X} + \boldsymbol{\Gamma}  \mathbf{K}  \boldsymbol{\Gamma}^{t} \right)^{-1}  \mathbf{X}^{t} \mathbf{X}  \left( \mathbf{X}^{t} \mathbf{X} + \boldsymbol{\Gamma}  \mathbf{K}  \boldsymbol{\Gamma}^{t} \right)^{-1}.
        \label{var_cov}
    \end{equation}

    Finally, by following \cite{Marquardt1970} (Theorem 8, page 594), the estimator given in expression (\ref{est_ridge}) is equivalent to the OLS estimator of the augmented model $\mathbf{Y}_{a} = \mathbf{X}_{a}  \boldsymbol{\beta} + \mathbf{u}_{a}$ where:
    \begin{equation}
        \label{matrices_aumentadas}
        \mathbf{Y}_{a} = \left(
            \begin{array}{c}
                \mathbf{Y} \\
                \mathbf{0}
            \end{array} \right), \quad
        \mathbf{X}_{a} = \left(
            \begin{array}{c}
                \mathbf{X} \\
                \mathbf{K}^{1/2}  \boldsymbol{\Gamma}^{t}
            \end{array} \right),
    \end{equation}
    where $\mathbf{0}$ is a vector of zeros with $p \times 1$ dimensions, due to $\widehat{\boldsymbol{\beta}}_{a} = \left( \mathbf{X}_{a}^{t} \mathbf{X}_{a} \right)^{-1} \mathbf{X}_{a}^{t} \mathbf{Y}_{a} = \left( \mathbf{X}^{t} \mathbf{X} + \boldsymbol{\Gamma}  \mathbf{K}  \boldsymbol{\Gamma}^{t} \right)^{-1} \mathbf{X}^{t} \mathbf{Y} = \widehat{\boldsymbol{\beta}}(\mathbf{K}).$

    However, this is the unique expression that coincides in the general ridge regression and the augmented model. Thus, for example, the matrix of variances and covariances of the augmented model is $var \left( \widehat{\boldsymbol{\beta}}_{a} \right) = \sigma^{2}  \left( \mathbf{X}_{a}^{t} \mathbf{X}_{a} \right)^{-1} = \sigma^{2}  \left( \mathbf{X}^{t} \mathbf{X} + \boldsymbol{\Gamma}  \mathbf{K}  \boldsymbol{\Gamma}^{t} \right)^{-1}$, which differs from the one obtained in (\ref{var_cov}), even when it is supposed that $\mathbf{u}$ and $\mathbf{u}_{a}$ present the same variance.

    \subsection{Particular cases}

\begin{itemize}
\item When $\mathbf{K} = k  \mathbf{I}$, the expressions obtained in this section coincide with those given by \cite{HoerlKennard1970b} and \cite{Marquardt1970}. In this case, the regular ridge (RR) uses the notation $\widehat{\boldsymbol{\beta}}(k)$ instead of $\widehat{\boldsymbol{\beta}}(\mathbf{K})$.
\item When $\mathbf{K} = diag(0,\dots,k_{l},\dots,0)$ it is verified that:
    $$\boldsymbol{\Gamma}  \boldsymbol{\Lambda}  \boldsymbol{\Gamma}^{t} = k_{l}  \left(
        \begin{array}{ccccc}
            \gamma_{1l}^{2} & \dots & \gamma_{1l}  \gamma_{ll} & \dots & \gamma_{1l}  \gamma_{pl} \\
            \vdots & \ddots & \vdots & \ddots & \vdots \\
            \gamma_{1l}  \gamma_{ll} & \dots & \gamma_{ll}^{2} & \dots & \gamma_{ll}  \gamma_{pl} \\
            \vdots & \ddots & \vdots & \ddots & \vdots \\
            \gamma_{1l}  \gamma_{pl} & \dots & \gamma_{pl}  \gamma_{ll} & \dots & \gamma_{pl}^{2}
        \end{array} \right).$$
    For the case $\mathbf{K} = k  \mathbf{I}$, only the elements of the main diagonal of matrix $\mathbf{X}^{t} \mathbf{X}$ are modified; and when $\mathbf{K} = diag(0,\dots,k_{l},\dots,0)$, all the elements of matrix $\mathbf{X}^{t} \mathbf{X}$ are modified. In this last case, the generalized ridge (GR), the notation $\widehat{\boldsymbol{\beta}}(k_{l})$ will be used instead of $\widehat{\boldsymbol{\beta}}(\mathbf{K})$.\\

    Finally, from expression (\ref{est_ridge_i}), it is obtained that:
    $$\widehat{\boldsymbol{\beta}}(k_{l})_{i} =  \frac{\gamma_{il}  \delta_{l}}{\lambda_{l}+k_{l}} + \sum \limits_{j=1,j\not=l}^{p} \frac{\gamma_{ij}  \delta_{j}}{\lambda_{j}}.$$
    As a consequence, $\lim \limits_{k_{l} \rightarrow +\infty} \widehat{\boldsymbol{\beta}}(k_{l})_{i} = \sum \limits_{j=1,j\not=l}^{p} \frac{\gamma_{ij}  \delta_{j}}{\lambda_{j}} = \widehat{\boldsymbol{\beta}}_{i} - \frac{\gamma_{il} \delta_{l}}{\lambda_{l}}$, where $\widehat{\boldsymbol{\beta}}_{i}$ is the element $i$ of $\widehat{\boldsymbol{\beta}}$. In other words, the estimations do not converge towards zero but around the OLS estimator.
\end{itemize}

\section{The Ridge Trace and Norm}
    \label{norm}

    In the work presented by \cite{HoerlKennard1970a}, where $\mathbf{K} = k  \mathbf{I}$, the trace of the ridge estimator is used to determine the values of $k$ that provide stable estimations. Thus, values of $\widehat{\boldsymbol{\beta}}(k)$ are represented as a function of a rank of values of $k$, usually $k \in [0, \ 1]$; and graphically is observed for what values of $k$ the estimations of $\widehat{\boldsymbol{\beta}}(k)$ are stabilized. \cite{HoerlKennard1970a} stated that \textit{coefficients chosen from a $k$ in this range will undoubtedly be closer to $\boldsymbol{\beta}$ and more stable for prediction than the least squares coefficients}.

    This way to select $k$ is justified by \cite{Marquardt1970} (Theorem 2, page 593) who shows that the norm of the ridge estimator, $|| \widehat{\boldsymbol{\beta}}(k) || = \widehat{\boldsymbol{\beta}}(k)^{t} \widehat{\boldsymbol{\beta}}(k)$, decreases when $k$ increases. In addition, for $k \rightarrow +\infty$, it is obtained that $|| \widehat{\boldsymbol{\beta}}(k) || \rightarrow 0$.

    This section analyzes the properties of  $|| \widehat{\boldsymbol{\beta}}(\mathbf{K}) ||$. Thus, considering (\ref{beta_K}), it is obtained that:
    \begin{equation}
        || \widehat{\boldsymbol{\beta}}(\mathbf{K}) || = \widehat{\boldsymbol{\beta}}(\mathbf{K})^{t} \widehat{\boldsymbol{\beta}}(\mathbf{K}) = \boldsymbol{\delta}^{t}  \boldsymbol{\Omega}  \boldsymbol{\Gamma}^{t}  \boldsymbol{\Gamma}  \boldsymbol{\Omega}  \boldsymbol{\delta} = \boldsymbol{\delta}^{t}  \boldsymbol{\Omega}^{2}  \boldsymbol{\delta} = \sum \limits_{j=1}^{p} \frac{\delta_{j}^{2}}{(\lambda_{j}+k_{j})^{2}}.
        \label{norma}
    \end{equation}
    It is evident that when $k_{j}$ ($j=1,\dots,p$) increases, $|| \widehat{\boldsymbol{\beta}}(\mathbf{K}) ||$ diminishes. Indeed, when $k_{j} \rightarrow +\infty$ for all $j$, it is verified that $|| \widehat{\boldsymbol{\beta}}(\mathbf{K}) || \rightarrow 0$.

    Consequently, a combination of values for $k_{1},\dots,k_{p}$ that allow stable estimations of the coefficients of the model can exist.
    \subsection{Particular cases}
\begin{itemize}
\item  When $\mathbf{K} = k  \mathbf{I}$, it is obtained that $|| \widehat{\boldsymbol{\beta}}(k) || = \sum \limits_{j=1}^{p} \frac{\delta_{j}^{2}}{(\lambda_{j}+k)^{2}}$. Thus, the indications of Theorem 2 of \cite{Marquardt1970} are clear.
\item When $\mathbf{K} = diag(0,\dots,k_{l},\dots,0)$:
    $$|| \widehat{\boldsymbol{\beta}}(k_{l}) || = \frac{\delta_{l}^{2}}{(\lambda_{l}+k_{l})^{2}} + \sum \limits_{j=1,j \not= l}^{p} \frac{\delta_{j}^{2}}{\lambda_{j}^{2}} = \frac{\delta_{l}^{2}}{(\lambda_{l}+k_{l})^{2}} + || \widehat{\boldsymbol{\beta}} || - \frac{\delta_{l}^{2}}{\lambda_{l}^{2}}
    \rightarrow \lim_{k_{l} \rightarrow +\infty} || \widehat{\boldsymbol{\beta}}(k_{l}) || = || \widehat{\boldsymbol{\beta}} || - \frac{\delta_{l}^{2}}{\lambda_{l}^{2}}.$$
    Thus, in this case, in addition to the possibility of obtaining estimations of $\boldsymbol{\beta}$ stable for some value of $k_{l}$, the norm of the estimator (\ref{est_ridge}) converges towards the norm of the OLS estimator.
\end{itemize}

\section{Mean Squared Error}
    \label{mse}

    Because the estimator $\widehat{\boldsymbol{\beta}}(\mathbf{K})$ given in (\ref{est_ridge}) is biased, it is interesting to calculate its mean squared error (MSE) and compare it to the one obtained from OLS.

    In this case, the MSE of $\widehat{\boldsymbol{\beta}}(\mathbf{K})$ will be given by:
    $$MSE \left( \widehat{\boldsymbol{\beta}}(\mathbf{K}) \right) = trace \left( var \left( \widehat{\boldsymbol{\beta}}(\mathbf{K}) \right) \right) + \left( E \left[ \widehat{\boldsymbol{\beta}}(\mathbf{K}) \right] - \boldsymbol{\beta} \right)^{t} \left( E \left[ \widehat{\boldsymbol{\beta}}(\mathbf{K}) \right] - \boldsymbol{\beta} \right) = \eta_{1}(\mathbf{K}) + \eta_{2}(\mathbf{K}).$$

    Due to $\left( \mathbf{X}^{t} \mathbf{X} + \boldsymbol{\Gamma}  \mathbf{K}  \boldsymbol{\Gamma}^{t} \right)^{-1} = \boldsymbol{\Gamma}  \boldsymbol{\Omega}  \boldsymbol{\Gamma}^{t}$ and $\mathbf{X}^{t} \mathbf{X} = \boldsymbol{\Gamma}  \boldsymbol{\Lambda}  \boldsymbol{\Gamma}^{t}$, from expression (\ref{var_cov}), the following is obtained:
    $$var \left( \widehat{\boldsymbol{\beta}}(\mathbf{K}) \right) = \sigma^{2}  \boldsymbol{\Gamma}  \boldsymbol{\Omega}  \boldsymbol{\Gamma}^{t}  \boldsymbol{\Gamma}  \boldsymbol{\Lambda}  \boldsymbol{\Gamma}^{t}  \boldsymbol{\Gamma}  \boldsymbol{\Omega}  \boldsymbol{\Gamma}^{t} = \sigma^{2}  \boldsymbol{\Gamma}  \boldsymbol{\Psi}  \boldsymbol{\Gamma}^{t},$$
    where $\boldsymbol{\Psi} = \boldsymbol{\Omega}  \boldsymbol{\Lambda}  \boldsymbol{\Omega} = diag \left( \frac{\lambda_{1}}{(\lambda_{1}+k_{1})^{2}}, \dots, \frac{\lambda_{p}}{(\lambda_{p}+k_{p})^{2}} \right)$. As consequence:
    \begin{eqnarray}
        \eta_{1}(\mathbf{K}) &=& trace \left( var \left( \widehat{\boldsymbol{\beta}}(\mathbf{K}) \right) \right) = \sigma^{2}  trace \left( \boldsymbol{\Gamma}  \boldsymbol{\Psi}  \boldsymbol{\Gamma}^{t} \right) = \sigma^{2}  trace \left( \boldsymbol{\Psi}  \boldsymbol{\Gamma}  \boldsymbol{\Gamma}^{t} \right) \nonumber \\
            &=& \sigma^{2}  trace \left( \boldsymbol{\Psi} \right) = \sigma^{2}  \sum \limits_{j=1}^{p} \frac{\lambda_{j}}{(\lambda_{j}+k_{j})^{2}}. \label{mse_1}
    \end{eqnarray}

    Furthermore, as $E \left[ \widehat{\boldsymbol{\beta}}(\mathbf{K}) \right] = \mathbf{W}_{K}  \boldsymbol{\beta}$ it is verified that:
    \begin{eqnarray}
        \eta_{2}(\mathbf{K}) &=& \left( E \left[ \widehat{\boldsymbol{\beta}}(\mathbf{K}) \right] - \boldsymbol{\beta} \right)^{t} \left( E \left[ \widehat{\boldsymbol{\beta}}(\mathbf{K}) \right] - \boldsymbol{\beta} \right) = \boldsymbol{\beta}^{t}  \left( \mathbf{W}_{K} - \mathbf{I} \right)^{t}  \left( \mathbf{W}_{K} - \mathbf{I} \right)  \boldsymbol{\beta} \nonumber \\
            &=& \boldsymbol{\xi}^{t}  \boldsymbol{\Theta}  \boldsymbol{\xi} = \sum \limits_{j=1}^{p} \frac{k_{j}^{2}  \xi_{j}^{2}}{(\lambda_{j}+k_{j})^{2}}, \label{mse_2}
    \end{eqnarray}
    where it was applied that:
    \begin{eqnarray*}
        \mathbf{W}_{K} &=& \boldsymbol{\Gamma}  \boldsymbol{\Omega}  \boldsymbol{\Gamma}^{t}  \boldsymbol{\Gamma}  \boldsymbol{\Lambda}  \boldsymbol{\Gamma}^{t}     \rightarrow \mathbf{W}_{K} - \mathbf{I} = \boldsymbol{\Gamma}  \left( \boldsymbol{\Omega}  \boldsymbol{\Lambda} - \mathbf{I} \right)  \boldsymbol{\Gamma}^{t} \\
            & & \rightarrow \left( \mathbf{W}_{K} - \mathbf{I} \right)^{t}  \left( \mathbf{W}_{K} - \mathbf{I} \right) = \boldsymbol{\Gamma}  \left( \boldsymbol{\Omega}  \boldsymbol{\Lambda} - \mathbf{I} \right)  \boldsymbol{\Gamma}^{t}  \boldsymbol{\Gamma}  \left( \boldsymbol{\Omega}  \boldsymbol{\Lambda} - \mathbf{I} \right)  \boldsymbol{\Gamma}^{t} \\
            & & \rightarrow \boldsymbol{\beta}^{t}  \left( \mathbf{W}_{K} - \mathbf{I} \right)^{t}  \left( \mathbf{W}_{K} - \mathbf{I} \right)  \boldsymbol{\beta} = \boldsymbol{\xi}^{t}  \boldsymbol{\Theta}  \boldsymbol{\xi},
    \end{eqnarray*}
    being $\boldsymbol{\Theta} = \left( \boldsymbol{\Omega}  \boldsymbol{\Lambda} - \mathbf{I} \right)  \left( \boldsymbol{\Omega}  \boldsymbol{\Lambda} - \mathbf{I} \right) = diag \left( \frac{k_{1}^{2}}{(\lambda_{1}+k_{1})^{2}}, \dots, \frac{k_{p}^{2}}{(\lambda_{p}+k_{p})^{2}} \right)$.

    As a consequence:
    \begin{equation}
        \label{MSE}
        MSE \left( \widehat{\boldsymbol{\beta}}(\mathbf{K}) \right) = \sum \limits_{j=1}^{p} \frac{\sigma^{2}  \lambda_{j} + k_{j}^{2}  \xi_{j}^{2}}{(\lambda_{j} + k_{j})^{2}}.
    \end{equation}
    It can be noted that when $k_{j} \rightarrow +\infty$ for all $j$, it is obtained that $MSE \left( \widehat{\boldsymbol{\beta}}(\mathbf{K}) \right) \rightarrow || \boldsymbol{\beta} ||$ due to:
    $$\lim \limits_{k_{j} \rightarrow + \infty} \eta_{1}(\mathbf{K}) = 0, \quad
    \lim \limits_{k_{j} \rightarrow + \infty} \eta_{1}(\mathbf{K}) = \sum \limits_{j=1}^{p} \xi_{j}^{2} = \boldsymbol{\xi}^{t}  \boldsymbol{\xi} = \boldsymbol{\beta}^{t}  \boldsymbol{\Gamma}  \boldsymbol{\Gamma}^{t}  \boldsymbol{\beta} = \boldsymbol{\beta}^{t}  \boldsymbol{\beta} = || \boldsymbol{\beta} ||.$$

    \subsection{Particular cases}

\begin{itemize}
    \item When $\mathbf{K} = k  \mathbf{I}$, the results are the same that those obtained by \cite{HoerlKennard1970b}. Thus, the expression of the MSE is given by:
    \begin{equation}
        \label{mse_ols}
        MSE \left( \widehat{\boldsymbol{\beta}}(k) \right) = \sum \limits_{j=1}^{p} \frac{\sigma^{2}  \lambda_{j} + k^{2}  \xi_{j}^{2}}{(\lambda_{j} + k)^{2}},
    \end{equation}
    $\eta_{1}(k) = \sigma^{2}  \sum \limits_{j=1}^{p} \frac{\lambda_{j}}{(\lambda_{j}+k)^{2}}$ is a continuous and monotonically decreasing function of $k$ while $\eta_{2}(k) = k^{2}  \sum \limits_{j=1}^{p} \frac{\xi_{j}}{(\lambda_{j}+k)^{2}}$ is a continuous and monotonically increasing function of $k$. In addition, $MSE \left( \widehat{\boldsymbol{\beta}}(k) \right) < MSE \left( \widehat{\boldsymbol{\beta}} \right)$ if $k < \sigma^{2}/\xi_{max}^{2}$, where $\xi_{max}$ is the maximum value of $\boldsymbol{\xi}$ and $MSE \left( \widehat{\boldsymbol{\beta}} \right) = \sigma^{2} \sum \limits_{j=1}^{p} \frac{1}{\lambda_{j}}$ is the MSE for the OLS estimator of model (\ref{modelo}).

    \item When $\mathbf{K} = diag(0,\dots,k_{l},\dots,0)$:
    \begin{equation}
        \label{mse_kl}
        MSE \left( \widehat{\boldsymbol{\beta}}(k_{l}) \right) = \frac{\sigma^{2}  \lambda_{l} + k_{l}^{2}  \xi_{l}^{2}}{(\lambda_{l}+k_{l})^{2}} + \sigma^{2}  \sum \limits_{j=1,j\not=l} \frac{1}{\lambda_{j}} = \frac{\sigma^{2}  \lambda_{l} + k_{l}^{2}  \xi_{l}^{2}}{(\lambda_{l}+k_{l})^{2}} + MSE \left( \boldsymbol{\beta} \right) - \frac{\sigma^{2}}{\lambda_{l}},
    \end{equation}
    and, consequently,
    \begin{eqnarray*}
        \frac{\partial MSE \left( \widehat{\boldsymbol{\beta}}(k_{l}) \right)}{\partial k_{l}} &=& \frac{2 k_{l} \xi_{l}^{2} (\lambda_{l}+k_{l})^{2} - (\sigma^{2} \lambda_{l}+k_{l}^{2}\xi_{l}^{2}) 2 (\lambda_{l}+k_{l})}{(\lambda_{l}+k_{l})^{2}} = \frac{2 (\lambda_{l}+k_{l}) \left( k_{l} \xi_{l}^{2} (\lambda_{l}+k_{l}) - \sigma^{2} \lambda_{l} - k_{l}^{2} \xi_{l}^{2} \right)}{(\lambda_{l}+k_{l})^{2}} \\
            &=& \frac{2 (\lambda_{l}+k_{l}) \lambda_{l} \left( k_{l} \xi_{l}^{2} - \sigma^{2} \right)}{(\lambda_{l}+k_{l})^{2}}.
    \end{eqnarray*}
    In that case, due to $\lambda_{l}, k_{l} > 0$:
    $$\frac{\partial MSE \left( \widehat{\boldsymbol{\beta}}(k_{l}) \right)}{\partial k_{l}} = 0 \leftrightarrow  k_{l} \xi_{l}^{2} - \sigma^{2} = 0 \leftrightarrow k_{l} = \frac{\sigma^{2}}{\xi_{l}^{2}}.$$

    Additionally, the particular point $k_{l,min} = \sigma^{2}/\xi_{l}^{2}$ is a minimum due to:
    \begin{eqnarray*}
        \frac{\partial^{2} MSE \left( \widehat{\boldsymbol{\beta}}(k_{l}) \right)}{\partial k_{l}^{2}} &=& 2  \frac{(\xi_{l}^{2} \lambda_{l}^{2} + 2k_{l} \xi_{l}^{2} \lambda_{l} - \sigma^{2} \lambda_{l})(\lambda_{l}+k_{l})^{2} - (k_{l} \xi_{l}^{2} \lambda_{l}^{2} - \sigma^{2} \lambda_{l}^{2} + k_{l}^{2} \xi_{l}^{2} \lambda_{l} - \sigma^{2} \lambda_{l} k_{l}) 2 (\lambda_{l}+k_{l}))}{(\lambda_{l}+k_{l})^{4}} \\
            &=& \frac{2}{(\lambda_{l}+k_{l})^{3}}  (\xi_{l}^{2} \lambda_{l}^{3} + \sigma^{2} \lambda_{l}^{2} + \xi_{l}^{2} \lambda_{l}^{2} k_{l} + \sigma^{2} \lambda_{l} k_{l} ) >0.
    \end{eqnarray*}

    Furthermore, it is verified that:
    $$MSE \left( \widehat{\boldsymbol{\beta}}(k_{l,min}) \right) = MSE \left( \widehat{\boldsymbol{\beta}} \right) - \frac{\sigma^{2}}{\lambda_{l}} + \frac{\sigma^{2} \xi_{l}^{4} \lambda_{l} + \sigma^{4} \xi_{l}^{2}}{(\xi_{l}^{2} \lambda_{l} + \sigma^{2})^{2}},$$
    and, consequently,  $MSE \left( \widehat{\boldsymbol{\beta}}(k_{l,min}) \right) < MSE \left( \widehat{\boldsymbol{\beta}} \right)$ if
    $$\frac{\xi_{l}^{4} \lambda_{l} + \sigma^{2} \xi_{l}^{2}}{(\xi_{l}^{2} \lambda_{l} + \sigma^{2})^{2}} - \frac{1}{\lambda_{l}} < 0
    \leftrightarrow
    \frac{\xi_{l}^{4} \lambda_{l}^{2} + \xi_{l}^{2} \lambda_{l} \sigma^{2} - (\xi_{l}^{2} \lambda_{l} + \sigma^{2})^{2}}{(\xi_{l}^{2}+ \lambda_{l} + \sigma^{2})^{2} \lambda_{l}} <0
    \leftrightarrow
    - \xi_{l}^{2} \lambda_{l} \sigma^{2} - \sigma^{4} < 0,$$
    which is true since $\lambda_{l} >0$. Then, for $\mathbf{K} = diag(0,\dots,k_{l,min},\dots,0)$, the estimator given in (\ref{est_ridge}) presents a lower MSE than the one obtained from the OLS estimator.

    Finally, considering expression (\ref{mse_kl}), it is verified that:
    $$\lim \limits_{k_{l} \rightarrow +\infty} MSE \left( \widehat{\boldsymbol{\beta}}(k_{l}) \right) = \xi_{l}^{2} - \frac{\sigma^{2}}{\lambda_{l}} + MSE \left( \widehat{\boldsymbol{\beta}}\right).$$
    Then, $\lim \limits_{k_{l} \rightarrow +\infty} MSE \left( \widehat{\boldsymbol{\beta}}(k_{l}) \right) < MSE \left( \widehat{\boldsymbol{\beta}}\right)$ if $\xi_{l}^{2} - \frac{\sigma^{2}}{\lambda_{l}} <0$. %Teniendo en cuenta que $k_{l,min} = \sigma^{2}/\xi_{l}^{2}$, la condición anterior es equivalente a que $k_{l,min} > \lambda_{l}$.

    Because $MSE \left( \widehat{\boldsymbol{\beta}}(k_{l}) \right)$ is increasing from $k_{l,min}$ (its derivative is positive for $k_{l,min} < k_{l}$) and decreasing before $k_{l,min}$ (its derivative is negative for $k_{l} < k_{l,min}$) and is a convex function (its second derivative is always positive), Figures \ref{graf_mse_1} and \ref{graf_mse_2} show the graphical representation of the MSE depending on whether the difference $\xi_{l}^{2} - \frac{\sigma^{2}}{\lambda_{l}}$ is negative or positive. Note that in the first case, $MSE \left( \widehat{\boldsymbol{\beta}}(k_{l}) \right)$ is always lower than $MSE \left( \widehat{\boldsymbol{\beta}} \right)$ regardless of the value of $k_{l}$.
\end{itemize}

\begin{figure}
    \begin{center}
        %\begin{minipage}[t]{6cm}
        {\scriptsize
        \begin{tikzpicture}
        % horizontal axis
        \draw[->] (0,0) -- (6,0) node[anchor=north] {$k_{l}$}; % \draw[->] no me funciona
        % vertical axis
        \draw[->] (0,0) -- (0,4);
        \draw (0,3) .. controls (0.9,0.5) and (1.2,0.5) .. (2.5,1.75);
%        \draw (0,2.5) .. controls (1,0.5) .. (2,0.5);
%        \draw (2,0.5) .. controls (3,2.75) .. (3.5,3.75);
        \draw (2.5,1.75) to[out=46,in=180] (5.5,2.43);
        % información eje
        %\foreach \y/\ytext in {1/1, 2/2, 3/3}
        %\draw[shift={(0,\y)}] (0pt,2pt) -- (0pt,-2pt) node[below] {$\ytext$};
        \draw[fill=black]
                    (0.6,4.3) node[left] {$MSE(\widehat{\boldsymbol{\beta}} (k_{l}))$}
                    (0,3) node[left] {$MSE(\widehat{\boldsymbol{\beta}})$}
                    (0,2.5) node[left] {$\lim_{k_{l} \rightarrow +\infty} MSE(\widehat{\boldsymbol{\beta}} (k_{l}))$}
                    %(0,2.5) node[left] {$MSE(\hat{\beta} (0))$}
                    %(0,1.75) node[left] {$MSE(\hat{\beta} (0.5+1.5h))$}
                    (0,1) node[left] {$MSE(\widehat{\boldsymbol{\beta}} (k_{l,min}))$}
                    (1.75,-0.3) node[left] {$k_{l,min}$}
                    %(3.15,-0.3) node[left] {$0.5+1.5h$}
                    (0,0) node[left] {$0$};
                    %(0,1.25) node[right] {No tiene sentido};
        % linea horizontal discontinua
        \draw[dotted] (0,3) -- (6,3);
        \draw[dotted] (0,2.5) -- (6,2.5);
        \draw (0,3) node {$\bullet$};
        \draw[dashed] (0,0.9) -- (1.2,0.9);
        \draw[dashed] (1.2,0) -- (1.2,0.9);
        \draw (1.2,0.9) node {$\bullet$};
        \end{tikzpicture}}
        %\end{minipage}
        \caption{$MSE \left( \widehat{\boldsymbol{\beta}}(k_{l}) \right)$ representation for $\xi_{l}^{2} - \frac{\sigma^{2}}{\lambda_{l}} < 0$}
        \label{graf_mse_1}
    \end{center}
\end{figure}
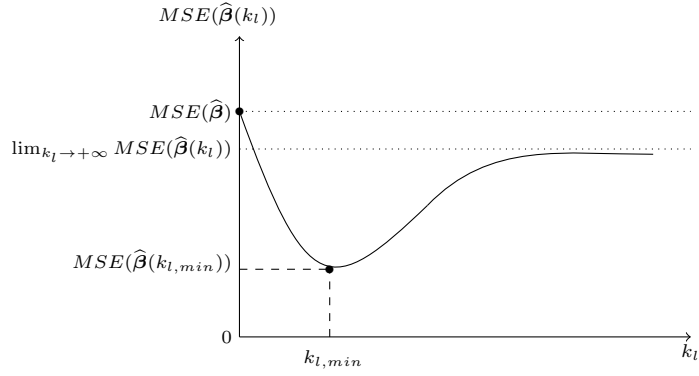

\begin{figure}
        \begin{center}
        {\scriptsize
            \begin{tikzpicture}
                % horizontal axis
                    \draw[->] (0,0) -- (6,0) node[anchor=north] {$k_{l}$}; % \draw[->] no me funciona
                % vertical axis
                    \draw[->] (0,0) -- (0,4);
                    \draw (0,3) .. controls (0.9,0.5) and (1.2,0.5) .. (2.5,1.75);
            %        \draw (0,2.5) .. controls (1,0.5) .. (2,0.5);
            %        \draw (2,0.5) .. controls (3,2.75) .. (3.5,3.75);
                    \draw (2.5,1.75) to[out=46,in=180] (5.5,3.45);
                % información eje
                %\foreach \y/\ytext in {1/1, 2/2, 3/3}
                %\draw[shift={(0,\y)}] (0pt,2pt) -- (0pt,-2pt) node[below] {$\ytext$};
                \draw[fill=black]
                    (0.6,4.3) node[left] {$MSE(\widehat{\boldsymbol{\beta}} (k_{l}))$}
                    (0,3) node[left] {$MSE(\widehat{\boldsymbol{\beta}})$}
                    (0,3.5) node[left] {$\lim_{k_{l} \rightarrow +\infty} MSE(\widehat{\boldsymbol{\beta}} (k_{l}))$}
                    %(0,2.5) node[left] {$MSE(\hat{\beta} (0))$}
                    %(0,1.75) node[left] {$MSE(\hat{\beta} (0.5+1.5h))$}
                    (0,1) node[left] {$MSE(\widehat{\boldsymbol{\beta}} (k_{l,min}))$}
                    (1.75,-0.3) node[left] {$k_{l,min}$}
                    %(3.15,-0.3) node[left] {$0.5+1.5h$}
                    (0,0) node[left] {$0$};
                    %(0,1.25) node[right] {No tiene sentido};
                % linea horizontal discontinua
                \draw[dotted] (0,3) -- (6,3);
                \draw[dotted] (0,3.5) -- (6,3.5);
                \draw (0,3) node {$\bullet$};
                \draw[dashed] (0,0.9) -- (1.2,0.9);
                \draw[dashed] (1.2,0) -- (1.2,0.9);
                \draw (1.2,0.9) node {$\bullet$};
            \end{tikzpicture}
        }
        \end{center}
        \caption{$MSE \left( \widehat{\boldsymbol{\beta}}(k_{l}) \right)$ representation for $\xi_{l}^{2} - \frac{\sigma^{2}}{\lambda_{l}} > 0$}\label{graf_mse_2}
\end{figure}
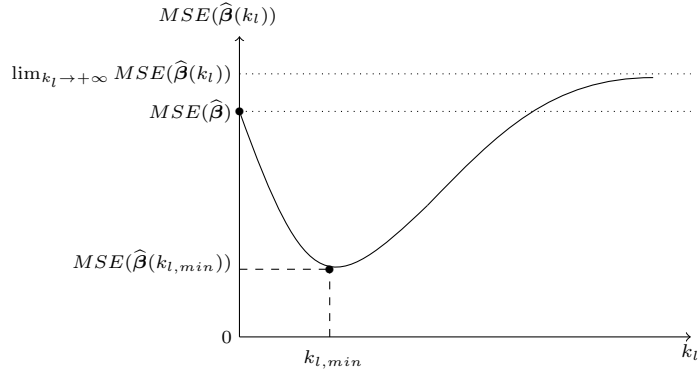

\section{Goodness of fit}\label{R2}

Although cross-validation techniques are often used to analyze the goodness of fit of the performed ridge estimation, this section proposes a goodness of fit measure that is the natural extension of the one used in OLS. Thus, given a model similar to (\ref{modelo}) with or without an intercept, it is verified\footnote{
        In fact, $\mathbf{Y}^{t} \mathbf{Y} = (\widehat{\mathbf{Y}} + \mathbf{e})^{t}(\widehat{\mathbf{Y}} + \mathbf{e}) = \widehat{\mathbf{Y}}^{t} \widehat{\mathbf{Y}} + \widehat{\mathbf{Y}}^{t} \mathbf{e} + \mathbf{e}^{t} \widehat{\mathbf{Y}} + \mathbf{e}^{t} \mathbf{e} = \widehat{\mathbf{Y}}^{t} \widehat{\mathbf{Y}} + \mathbf{e}^{t} \mathbf{e}$, where it was considered that $\mathbf{e}^{t} \widehat{\mathbf{Y}} = \widehat{\mathbf{Y}}^{t} \mathbf{e} = \widehat{\boldsymbol{\beta}}^{t} \mathbf{X}^{t} \mathbf{e} = 0$.
    }
    the decomposition $\mathbf{Y}^{t} \mathbf{Y} = \widehat{\mathbf{Y}}^{t}\widehat{\mathbf{Y}} + \mathbf{e}^{t} \mathbf{e}$, where $\mathbf{e} = \mathbf{Y} - \widehat{\mathbf{Y}}$ are the residuals of such a model.

   In this case, the goodness of fit is defined as:
    $$GoF = \frac{\widehat{\mathbf{Y}}^{t} \widehat{\mathbf{Y}}}{\mathbf{Y}^{t} \mathbf{Y}} = 1 - \frac{\mathbf{e}^{t} \mathbf{e}}{\mathbf{Y}^{t} \mathbf{Y}}.$$
    Appendix \ref{appendixGoF} shows that this measure is affected by origin changes but not by scale changes. A particular interesting case is when the dependent variable presents zero mean because the decomposition $\mathbf{Y}^{t} \mathbf{Y} = \widehat{\mathbf{Y}}^{t}\widehat{\mathbf{Y}} + \mathbf{e}^{t} \mathbf{e}$ coincides with the sum of squares decomposition\footnote{
        As shown in \cite{Rodriguez2019R2}, this decomposition of the sum of squares is not verified in the ridge estimation and, consequently, cannot be applied to define a measure of its goodness of fit.} traditionally applied to calculate the coefficient of determination, $R^{2}$. That is, if $\overline{\mathbf{Y}} = 0$, then $GoF = R^{2}$ (see \cite{Salmeron2020centered} for more details).

    Defining the residuals of the ridge regression as $\mathbf{e} (\mathbf{K}) = \mathbf{Y} - \widehat{\mathbf{Y}}(\mathbf{K}) = \mathbf{Y} - \mathbf{X}  \widehat{\boldsymbol{\beta}}(\mathbf{K})$, where $\widehat{\boldsymbol{\beta}}(\mathbf{K})$ is given in expression (\ref{est_ridge}), it is obtained that $\mathbf{Y}^{t} \mathbf{Y} = \widehat{\mathbf{Y}}(\mathbf{K})^{t}\widehat{\mathbf{Y}}(\mathbf{K}) + 2 \widehat{\mathbf{Y}}(\mathbf{K})^{t} \mathbf{e}(\mathbf{K}) + \mathbf{e}(\mathbf{K})^{t} \mathbf{e}(\mathbf{K})$.
    Since:
    \begin{eqnarray*}
        \widehat{\mathbf{Y}}(\mathbf{K})^{t} \mathbf{e}(\mathbf{K}) &=& \widehat{\boldsymbol{\beta}}(\mathbf{K})^{t} \mathbf{X}^{t}  \left( \mathbf{Y} - \mathbf{X}  \widehat{\boldsymbol{\beta}}(\mathbf{K}) \right) = \widehat{\boldsymbol{\beta}}(\mathbf{K})^{t}  \left( \mathbf{X}^{t} \mathbf{Y} - \mathbf{X}^{t} \mathbf{X}  \widehat{\boldsymbol{\beta}}(\mathbf{K}) \right) \\
            &=& \widehat{\boldsymbol{\beta}}(\mathbf{K})^{t}  \left( \mathbf{X}^{t} \mathbf{X} + \boldsymbol{\Gamma} \mathbf{K} \boldsymbol{\Gamma}^{t} - \mathbf{X}^{t} \mathbf{X} \right)  \widehat{\boldsymbol{\beta}}(\mathbf{K}) = \widehat{\boldsymbol{\beta}}(\mathbf{K})^{t}  \boldsymbol{\Gamma} \mathbf{K} \boldsymbol{\Gamma}^{t}  \widehat{\boldsymbol{\beta}}(\mathbf{K}) \\
        \widehat{\mathbf{Y}}(\mathbf{K})^{t}\widehat{\mathbf{Y}}(\mathbf{K}) &=& \widehat{\boldsymbol{\beta}}(\mathbf{K})^{t}  \mathbf{X}^{t} \mathbf{X}  \widehat{\boldsymbol{\beta}}(\mathbf{K}),
    \end{eqnarray*}
    it is verified that:
    \begin{equation}
        \label{sumas_K}
        \mathbf{Y}^{t} \mathbf{Y} = \widehat{\boldsymbol{\beta}}(\mathbf{K})^{t}  \left( \mathbf{X}^{t} \mathbf{X} + 2 \boldsymbol{\Gamma} \mathbf{K} \boldsymbol{\Gamma}^{t} \right) \widehat{\boldsymbol{\beta}}(\mathbf{K}) + \mathbf{e}(\mathbf{K})^{t} \mathbf{e}(\mathbf{K}),
    \end{equation}
    where $\mathbf{Y}^{t} \mathbf{Y}$ is the total sum of squares, $\mathbf{e}(\mathbf{K})^{t} \mathbf{e}(\mathbf{K})$ is the residual sum of squares and $\widehat{\boldsymbol{\beta}}(\mathbf{K})^{t}  \left( \mathbf{X}^{t} \mathbf{X} + 2 \boldsymbol{\Gamma} \mathbf{K} \boldsymbol{\Gamma}^{t} \right) \widehat{\boldsymbol{\beta}}(\mathbf{K})$ is identified with the explained sum of squares of the generalized ridge regression.

    In this case, the goodness of fit of the ridge estimation can be defined with the following expression:
    \begin{equation}
        \label{ba_K}
        GoF (\mathbf{K}) = \frac{\widehat{\boldsymbol{\beta}}(\mathbf{K})^{t}  \left( \mathbf{X}^{t} \mathbf{X} + 2 \boldsymbol{\Gamma} \mathbf{K} \boldsymbol{\Gamma}^{t} \right) \widehat{\boldsymbol{\beta}}(\mathbf{K})}{\mathbf{Y}^{t} \mathbf{Y}} = 1 - \frac{\mathbf{e}(\mathbf{K})^{t} \mathbf{e}(\mathbf{K})}{\mathbf{Y}^{t} \mathbf{Y}}.
    \end{equation}

    Considering that $\widehat{\boldsymbol{\beta}}(\mathbf{K}) = \boldsymbol{\Gamma}  \boldsymbol{\Omega}  \boldsymbol{\delta}$ and $\mathbf{X}^{t} \mathbf{X} = \boldsymbol{\Gamma} \boldsymbol{\Lambda} \boldsymbol{\Gamma}^{t}$, it is obtained that:
    \begin{eqnarray*}
        \widehat{\boldsymbol{\beta}}(\mathbf{K})^{t}  \left( \mathbf{X}^{t} \mathbf{X} + 2  \boldsymbol{\Gamma}  \mathbf{K}  \boldsymbol{\Gamma}^{t} \right) \widehat{\boldsymbol{\beta}}(\mathbf{K}) &=& \boldsymbol{\delta}^{t}  \boldsymbol{\Omega}  \boldsymbol{\Gamma}^{t} \left( \boldsymbol{\Gamma}  \boldsymbol{\Lambda}  \boldsymbol{\Gamma}^{t} + 2  \boldsymbol{\Gamma}  \mathbf{K}  \boldsymbol{\Gamma}^{t} \right)  \boldsymbol{\Gamma}  \boldsymbol{\Omega}  \boldsymbol{\delta} \\
            &=& \boldsymbol{\delta}^{t}  \boldsymbol{\Omega}  \left( \boldsymbol{\Lambda} + 2  \mathbf{K} \right)  \boldsymbol{\Omega}  \boldsymbol{\delta} = \boldsymbol{\delta}^{t}  \boldsymbol{\Xi} \boldsymbol{\delta},
    \end{eqnarray*}
    where $\Xi = diag \left( \frac{\lambda_{1} + 2 k_{1}}{(\lambda_{1} + k_{1})^{2}}, \dots, \frac{\lambda_{p} + 2 k_{p}}{(\lambda_{p} + k_{p})^{2}} \right)$.
    Then, the expression (\ref{ba_K}) can be given by:
    \begin{equation}
        \label{ba_K_bis}
        GoF (\mathbf{K}) = \frac{1}{\mathbf{Y}^{t} \mathbf{Y}}  \sum \limits_{j=1}^{p} \frac{\delta_{j}^{2}  (\lambda_{j} + 2 k_{j})}{(\lambda_{j} + k_{j})^{2}},
    \end{equation}
    and, consequently, $GoF (\mathbf{K}) \rightarrow 0$ when $k_{j} \rightarrow +\infty$ for all $j$.

    Finally, given the augmented model defined by the matrices given in (\ref{matrices_aumentadas}), it is verified that $\mathbf{Y}_{a}^{t} \mathbf{Y}_{a} = \widehat{\mathbf{Y}}_{a}^{t} \widehat{\mathbf{Y}}_{a} + \mathbf{e}_{a}^{t} \mathbf{e}_{a}$, where $\mathbf{e}_{a} = \mathbf{Y}_{a} - \widehat{\mathbf{Y}}_{a}$ are the residuals of that model, since:
    $$\widehat{\mathbf{Y}}_{a}^{t} \mathbf{e}_{a} = \widehat{\boldsymbol{\beta}}(\mathbf{K})^{t} \mathbf{X}_{a}^{t} \left( \mathbf{Y}_{a} - \mathbf{X}_{a} \widehat{\boldsymbol{\beta}}(\mathbf{K}) \right) = \widehat{\boldsymbol{\beta}}(\mathbf{K})^{t} \left( \mathbf{X}_{a}^{t}\mathbf{X}_{a} - \mathbf{X}_{a}^{t}\mathbf{X}_{a} \right) \widehat{\boldsymbol{\beta}}(\mathbf{K}) = 0.$$
    In this case, the goodness of fit can be defined as:
    \begin{equation}
        \label{R2_a}
        GoF_{a}(\mathbf{K}) = \frac{\widehat{\mathbf{Y}}_{a}^{t} \widehat{\mathbf{Y}}_{a}}{\mathbf{Y}_{a}^{t} \mathbf{Y}_{a}} = \frac{\widehat{\boldsymbol{\beta}}(\mathbf{K})^{t} \mathbf{X}_{a}^{t} \mathbf{X}_{a} \widehat{\boldsymbol{\beta}}(\mathbf{K})}{\mathbf{Y}^{t} \mathbf{Y}} = \frac{\widehat{\boldsymbol{\beta}}(\mathbf{K})^{t} \left( \mathbf{X}^{t} \mathbf{X} + \boldsymbol{\Gamma} \mathbf{K} \boldsymbol{\Gamma}^{t} \right) \widehat{\boldsymbol{\beta}}(\mathbf{K})}{\mathbf{Y}^{t} \mathbf{Y}}.
    \end{equation}
    Note that expressions (\ref{ba_K}) and (\ref{R2_a}) are slightly different.

    \subsection{Particular cases}
\begin{itemize}
\item When $\mathbf{K} = k  \mathbf{I}$, expression (\ref{ba_K}) can be expressed as:
    \begin{equation}
        \label{ba_k}
        GoF (k) = \frac{\widehat{\boldsymbol{\beta}}(k)^{t}  \left( \mathbf{X}^{t} \mathbf{X} + 2  k  \mathbf{I} \right) \widehat{\boldsymbol{\beta}}(k)}{\mathbf{Y}^{t} \mathbf{Y}} = 1 - \frac{\mathbf{e}(k)^{t} \mathbf{e}(k)}{\mathbf{Y}^{t} \mathbf{Y}},
    \end{equation}
    where $\mathbf{e}(k) = \mathbf{Y} - \mathbf{X}  \widehat{\boldsymbol{\beta}}(k)$.
    Analogously, expression (\ref{ba_K_bis}) can be rewritten as:
    $$GoF (k) = \frac{1}{\mathbf{Y}^{t} \mathbf{Y}}  \sum \limits_{j=1}^{p} \frac{\delta_{j}^{2}  (\lambda_{j} + 2 k)}{(\lambda_{j} + k)^{2}},$$
    and, then:
    $$\frac{\partial GoF (k)}{\partial k} = - \frac{1}{\mathbf{Y}^{t} \mathbf{Y}}  \sum \limits_{j=1}^{p} \frac{2 \delta_{j}^{2} (\lambda_{j}+k) k}{(\lambda_{j}+k)^{4}} < 0,$$
    i.e., $GoF(k)$ is decreasing as a function of $k$.
    In addition, $\lim \limits_{k \rightarrow +\infty} GoF(k) = 0$.

    Furthermore, \cite{Rodriguez2019R2} analyzed the coefficient of determination in the ridge regression, establishing that for a correct behavior of this measure, the data should be standardized and proposed (Theorem 4) the following expression:
    \begin{equation}
        \label{R2_ainara}
        GoF(k) = \widehat{\boldsymbol{\beta}}(k)^{t} \mathbf{x}^{t} \mathbf{y} + k  \widehat{\boldsymbol{\beta}}(k)^{t} \widehat{\boldsymbol{\beta}}(k),
    \end{equation}
   where $\mathbf{x}$ and $\mathbf{y}$ are the standardized versions of $\mathbf{X}$ and $\mathbf{Y}$, respectively. This measure decrease as a function of $k$.

    Note that for the case of standardized data, $\mathbf{y}^{t} \mathbf{y} = 1$ expressions (\ref{ba_k}) and (\ref{R2_ainara}) coincide due to:
    $$\widehat{\boldsymbol{\beta}}(k)^{t}  \left( \mathbf{x}^{t} \mathbf{x} + 2  k  \mathbf{I} \right) \widehat{\boldsymbol{\beta}}(k) = \widehat{\boldsymbol{\beta}}(k)^{t}  \left( \mathbf{x}^{t} \mathbf{x} + k  \mathbf{I} \right) \widehat{\boldsymbol{\beta}}(k) + k  \widehat{\boldsymbol{\beta}}(k)^{t} \widehat{\boldsymbol{\beta}}(k) = \widehat{\boldsymbol{\beta}}(k)^{t}  \mathbf{x}^{t} \mathbf{y} + k  \widehat{\boldsymbol{\beta}}(k)^{t} \widehat{\boldsymbol{\beta}}(k).$$

\item When $\mathbf{K} = diag(0,\dots,k_{l},\dots,0)$, from expression (\ref{ba_K_bis}), it is obtained that:
    \begin{equation}
        \label{ba_k_l}
        GoF(k_{l}) = \frac{1}{\mathbf{Y}^{t} \mathbf{Y}}  \left( \frac{\delta_{l}^{2}  (\lambda_{l}+2 k_{l})}{(\lambda_{l}+k_{l})^{2}} + \sum \limits_{j=1,j\not=l}^{p} \frac{\delta_{j}^{2}}{\lambda_{j}} \right).
    \end{equation}
   In that case:
    $$\frac{\partial GoF (k_{l})}{\partial k_{l}} = - \frac{1}{\mathbf{Y}^{t} \mathbf{Y}}  \frac{2 \delta_{l}^{2} (\lambda_{l}+k_{l}) k_{l}}{(\lambda_{l}+k_{l})^{4}} < 0,$$
    i.e., $GoF (k_{l})$ is a decreasing function in $k_{l}$. Finally:
    $$\lim \limits_{k_{l} \rightarrow + \infty} GoF(k_{l}) = \frac{1}{\mathbf{Y}^{t} \mathbf{Y}}  \left( \sum \limits_{j=1,j\not=l}^{p} \frac{\delta_{j}^{2}}{\lambda_{j}} \right) = GoF - \frac{1}{\mathbf{Y}^{t} \mathbf{Y}} \frac{\delta_{l}^{2}}{\lambda_{l}}.$$
\end{itemize}

\section{Comparison in terms of MSE criterion}
    \label{matrix_mse}

By following \cite{Theobald1974}, \cite{Farebrother1976}, \cite{Trenklar1980} and \cite{Salmeron2024}, it is possible to state the following result.

        \begin{proposition}
            Let $\widehat{\boldsymbol{\beta}}_{i} = \mathbf{C}_{i} \mathbf{Y}$, with $i=1,2$, be two linear estimators of $\boldsymbol{\beta}$ in equation (\ref{modelo}), if it is verified that $\mathbf{S} = \mathbf{C}_{2} \mathbf{C}_{2}^{t} - \mathbf{C}_{1} \mathbf{C}_{1}^{t}$ is a positive definite matrix, then the estimator $\widehat{\boldsymbol{\beta}}_{1}$ is better than estimator $\widehat{\boldsymbol{\beta}}_{2}$ under the root mean squared error matrix criterion and MSE criterion. That is, $\widehat{\boldsymbol{\beta}}_{1}$ is better than estimator $\widehat{\boldsymbol{\beta}}_{2}$ when the following inequality is verified:
            $$\boldsymbol{\beta}^{t} \left( \mathbf{C}_{1} \mathbf{X} - \mathbf{I} \right)^{t} \mathbf{S}^{-1} \left( \mathbf{C}_{1} \mathbf{X} - \mathbf{I} \right) \boldsymbol{\beta} < \sigma^{2},$$
            being $\mathbf{S}$ a positive definite matrix.
            \hfill $\lozenge$
        \end{proposition}

        Then, from the previous proposition it is possible to establish that $\widehat{\boldsymbol{\beta}}(\mathbf{K})$ with $\mathbf{K} = diag (k_{1}, k_{2}, \dots, k_{p})  \not= k \mathbf{I}$ is preferred over $\widehat{\boldsymbol{\beta}}(\mathbf{k})$ under the criterion of the matrix of the root mean squared if $k_{i} \geq k$ for all $i=1,\dots,p$.

        \begin{proposition}
            \label{proposition2}
            The generalized ridge estimator, $\widehat{\boldsymbol{\beta}}(\mathbf{K}) = \left( \mathbf{X}^{t} \mathbf{X} + \boldsymbol{\Gamma} \mathbf{K} \boldsymbol{\Gamma} \right)^{-1} \mathbf{X}^{t} \mathbf{Y}$ with $\mathbf{K}  = diag (k_{1}, k_{2}, \dots, k_{p}) \not= k \mathbf{I}$, is preferred over the regular ridge estimator, $\widehat{\boldsymbol{\beta}}(\mathbf{k}) = \left( \mathbf{X}^{t} \mathbf{X} + k \mathbf{I} \right)^{-1} \mathbf{X}^{t} \mathbf{Y}$, under the root mean squared error matrix criterion for values of $\mathbf{K}$ and $k$ that satisfy the following expression:
            $$\boldsymbol{\beta}^{t} \left( \left( \mathbf{X}^{t} \mathbf{X} + \boldsymbol{\Gamma} \mathbf{K} \boldsymbol{\Gamma} \right)^{-1} \mathbf{X}^{t} \mathbf{X} - \mathbf{I} \right)^{t} \mathbf{S}^{-1} \left( \left( \mathbf{X}^{t} \mathbf{X} + \boldsymbol{\Gamma} \mathbf{K} \boldsymbol{\Gamma} \right)^{-1} \mathbf{X}^{t} \mathbf{X} - \mathbf{I} \right) \boldsymbol{\beta} < \sigma^{2}.$$
            where $\mathbf{S} = \left( \left( \mathbf{X}^{t} \mathbf{X} + k \mathbf{I} \right)^{-1} - \left( \mathbf{X}^{t} \mathbf{X} + \boldsymbol{\Gamma} \mathbf{K} \boldsymbol{\Gamma} \right)^{-1} \right) \mathbf{X}^{t} \mathbf{X} \left( \left( \mathbf{X}^{t} \mathbf{X} + k \mathbf{I} \right)^{-1} - \left( \mathbf{X}^{t} \mathbf{X} + \boldsymbol{\Gamma} \mathbf{K} \boldsymbol{\Gamma} \right)^{-1} \right)$ is a positive definite matrix. \hfill $\lozenge$
        \end{proposition}

        \begin{proof}
            Considering $\mathbf{C}_{1} = \left( \mathbf{X}^{t} \mathbf{X} + \boldsymbol{\Gamma} \mathbf{K} \boldsymbol{\Gamma} \right)^{-1} \mathbf{X}^{t}$ and $\mathbf{C}_{2} =  \left( \mathbf{X}^{t} \mathbf{X} + k \mathbf{I} \right)^{-1} \mathbf{X}^{t}$, it is verified that:
            $$\mathbf{S} = \mathbf{C}_{2} \mathbf{C}_{2}^{t} - \mathbf{C}_{1} \mathbf{C}_{1}^{t} = \mathbf{A} \mathbf{X}^{t} \mathbf{X} \mathbf{A},$$
            where $\mathbf{A} = \left( \mathbf{X}^{t} \mathbf{X} + k \mathbf{I} \right)^{-1} - \left( \mathbf{X}^{t} \mathbf{X} + \boldsymbol{\Gamma} \mathbf{K} \boldsymbol{\Gamma} \right)^{-1}$.
            Then, if $\mathbf{A}$ is a positive definite matrix, as is $\mathbf{X}^{t} \mathbf{X}$, then their product, that is $\mathbf{S}$, is also a positive definite matrix.

            Taking into account that $\mathbf{X}^{t} \mathbf{X} = \boldsymbol{\Gamma} \boldsymbol{\Lambda} \boldsymbol{\Gamma}^{t}$ with $\boldsymbol{\Gamma} \boldsymbol{\Gamma}^{t} = \mathbf{I} = \boldsymbol{\Gamma}^{t} \boldsymbol{\Gamma}$:
            \begin{itemize}
                \item $\mathbf{X}^{t} \mathbf{X} + k \mathbf{I} = \boldsymbol{\Gamma} \boldsymbol{\Lambda} \boldsymbol{\Gamma}^{t} + k \boldsymbol{\Gamma} \boldsymbol{\Gamma}^{t}  = \boldsymbol{\Gamma} \mathbf{D}_{\lambda_{i}+k} \boldsymbol{\Gamma}^{t}$ where $\mathbf{D}_{\lambda_{i}+k}$ is a diagonal matrix with elements $\lambda_{i}+k$ for $i=1,\dots,p$.
                \item $\mathbf{X}^{t} \mathbf{X} + \boldsymbol{\Gamma} \mathbf{K} \boldsymbol{\Gamma} = \boldsymbol{\Gamma} \boldsymbol{\Lambda} \boldsymbol{\Gamma}^{t} + \boldsymbol{\Gamma} \mathbf{K} \boldsymbol{\Gamma} = \boldsymbol{\Gamma} \mathbf{D}_{\lambda_{i}+k_{i}} \boldsymbol{\Gamma}^{t}$ where $\mathbf{D}_{\lambda_{i}+k_{i}}$ is a diagonal matrix with elements $\lambda_{i}+k_{i}$ for $i=1,\dots,p$.
            \end{itemize}
            In that case, $\mathbf{A} = \boldsymbol{\Gamma} \left( \mathbf{D}_{\frac{1}{\lambda_{i}+k}} - \mathbf{D}_{\frac{1}{\lambda_{i}+k_{i}}} \right) \boldsymbol{\Gamma}^{t} = \boldsymbol{\Gamma} \mathbf{D}_{\frac{k_{i}-k}{(\lambda_{i}+k)(\lambda_{i}+k_{i})}}  \boldsymbol{\Gamma}^{t}$, and considering $\mathbf{a}_{p \times 1} = \left( a_{1}, a_{2}, \dots, a_{p} \right)^{t}$:
            $$\mathbf{a}^{t} \mathbf{A} \mathbf{a} = \mathbf{b}^{t} \mathbf{D}_{\frac{k_{i}-k}{(\lambda_{i}+k)(\lambda_{i}+k_{i})}} \mathbf{b} = \sum \limits_{i=1}^{p} \frac{(k_{i}-k) b_{i}^{2}}{(\lambda_{i}+k)(\lambda_{i}+k_{i})},$$
            where $\mathbf{b}_{p \times 1} = \boldsymbol{\Gamma}^{t} \mathbf{a}$.

            As $\lambda_{i} > 0$, $k \geq 0$ and $k_{i} \geq 0$ for all $i=1,\dots,p$, it is clear that $\mathbf{A}$ is a positive definite matrix if\footnote{
                Note that it is not possible that $k_{i} = k$ for all $i$, so there must exist an $i$ such that $k_{i} > k$.
            } $k_{i} \geq k$ for all $i=1,\dots,p$.
        \end{proof}

        Finally, from Proposition \ref{proposition2}, it is possible to state the following corollaries.

        \begin{corollary}
            \label{corolario1}
            The generalized ridge estimator, $\widehat{\boldsymbol{\beta}}(\mathbf{K}) = \left( \mathbf{X}^{t} \mathbf{X} + \boldsymbol{\Gamma} \mathbf{K} \boldsymbol{\Gamma} \right)^{-1} \mathbf{X}^{t} \mathbf{Y}$ with $\mathbf{K} = diag (k_{1}, k_{2}, \dots, k_{p}) \not= k \mathbf{I}$, is preferred over the OLS estimator, $\widehat{\boldsymbol{\beta}} = \left( \mathbf{X}^{t} \mathbf{X} \right)^{-1} \mathbf{X}^{t} \mathbf{Y}$, under the root mean squared error matrix criterion.
        \end{corollary}

        \begin{proof}
            Immediate as $k_{i} > 0$ for all $i=1,\dots,p$.
        \end{proof}

        \begin{corollary}
            \label{corolario2}
            The regular ridge estimator, $\widehat{\boldsymbol{\beta}}(\mathbf{k}) = \left( \mathbf{X}^{t} \mathbf{X} + k \mathbf{I} \right)^{-1} \mathbf{X}^{t} \mathbf{Y}$, is preferred over the OLS estimator, $\widehat{\boldsymbol{\beta}} = \left( \mathbf{X}^{t} \mathbf{X} \right)^{-1} \mathbf{X}^{t} \mathbf{Y}$, under the root mean squared error matrix criterion.
        \end{corollary}

        \begin{proof}
            Immediate as $k > 0$.
        \end{proof}

        \begin{corollary}
            \label{corolario3}
            The regular ridge estimator, $\widehat{\boldsymbol{\beta}}(\mathbf{k}) = \left( \mathbf{X}^{t} \mathbf{X} + k \mathbf{I} \right)^{-1} \mathbf{X}^{t} \mathbf{Y}$, is preferred over generalized ridge estimator, $\widehat{\boldsymbol{\beta}}(\mathbf{K}) = \left( \mathbf{X}^{t} \mathbf{X} + \boldsymbol{\Gamma} \mathbf{K} \boldsymbol{\Gamma} \right)^{-1} \mathbf{X}^{t} \mathbf{Y}$ with $\mathbf{K} = diag (0, \dots, k_{l}, \dots, 0)$, under the root mean squared error matrix criterion if $k > k_{l}$.
        \end{corollary}

        \begin{proof}
            Immediate taking into account that in this case:
            $$\mathbf{a}^{t} \mathbf{A} \mathbf{a} = \frac{(k - k_{l}) b_{l}^{2}}{(\lambda_{i} + k_{l})(\lambda_{i}+k)} +  \sum \limits_{i=1, i \not= l}^{p} \frac{k b_{i}^{2}}{\lambda_{i} (\lambda_{i} + k)} > 0,$$
            if $k > k_{l} > 0$.
        \end{proof}

\section{Inference}
    \label{boots}

        Although there have been various efforts to deal with inference in the ridge estimator (see, for example, \cite{Obenchain1975,obenchain1977classical}, \cite{halawa2000tests} or \cite{lmridge_paper}),  in this paper we will focus on the use of bootstrap methods (see, for example, Efron and Tibshirani \cite{EfronTibshirani1986}).

        Thus, given a fixed value of $\mathbf{K}$, obtained by any of the methods proposed in the previous subsections, the following steps will be performed:
        \begin{enumerate}[(i)]
            \item Generate randomly and with replacement $m$ subsamples of equal size to the original one. The value of $m$ must be large.
            \item For each previous subsample, the statistic of $\theta$ is calculated. Therefore, we have  $m$ values for that statistic: $\theta_{1},\dots,\theta_{m}$.
            \item Obtain the approximation of a confidence interval by the expression:
                $$[P_{0.025}(\theta_{1},\dots,\theta_{m}), P_{0.975}(\theta_{1},\dots,\theta_{m})],$$
                where the 0.025 and 0.975 percentiles of the $m$ values calculated in the second step have been considered as lower and upper extremes.
        \end{enumerate}
        The cases where $\theta$ equals $widehat{\boldsymbol{\beta}}(\mathbf{K})$ or $GoF(\mathbf{K})$ for the two particular cases analyzed are of interest in this paper.

\section{Example}
    \label{example}

       The contribution of this paper is illsutrated in this section with the example previously presented by \cite{HoerlKennard1970b}. We first present the results of sections \ref{estimation} to \ref{matrix_mse} and then we present the results of section \ref{boots}. The results obtained are compared with those provided by other R packages for the regular ridge estimator (such as, for example, \textit{lmridge} (\cite{lmridge}) and \textit{lrmest} (\cite{lrmest})). Note that the code in (\cite{RCoreTeam}) used to generate the results is available in Github, specifically at \url{https://github.com/rnoremlas/GRR/tree/main/01_Biased_estimation}.

    \subsection{Estimation properties, ridge trace, norm, mean squared error, goodness of fit and root mean squared error matrix criterion}

    To illustrate the contribution of this paper, this section uses the data set of \cite{GormanToman} also used by \cite{Rodriguez2019R2,Rodriguezetal2021} and \cite{HoerlKennard1970b}, who stated that \textit{Gorman and Toman use this problem as an example to portray a shortcut method for finding a ``best'' subset of factors of a specified size less than ten without having to compute all regressions of the specified size}. This dataset contains 11 independent variables; and contrary to Hoerl and Kennard, the intercept is considered.

    In this example, from expressions (\ref{beta_K}), (\ref{MSE}) and (\ref{ba_K}) we calculate the estimations, the mean squared error and the goodness of fit, respectively, for the following cases:
    \begin{enumerate}[a)]
        \item (OLS) $\mathbf{K} = diag(0,\dots,0)$;
        \item (RR) $\mathbf{K} = k \mathbf{I}$ being $k = p \cdot \frac{\sigma^{2}}{\boldsymbol{\beta}^{t} \boldsymbol{\beta}} = k_{HKB}$ (\cite{HoerlKennardBaldwin}), $k = \frac{\sigma^{2}}{\xi_{max}^{2}} = k_{HK}$ (\cite{HoerlKennard1970b}) and for the value $k = k_{min}$ the truly minimizes the $MSE \left( \widehat{\boldsymbol{\beta}}(k) \right)$ calculated (considering that the MSE first decreases but later increases) with the Algorithm \ref{algoritmo};
        \item (GR) $\mathbf{K} = diag(k_{1},\dots,k_{p})$ with $k_{i} = \frac{\sigma^{2}}{\xi_{i}^{2}}$ for $i=1,\dots,p$ (\cite{HoerlKennard1970b}) and
        \item (GR) $\mathbf{K} = diag(0,\dots,k_{l,min},\dots,0)$.
    \end{enumerate}
    For all these calculations, the estimation of $\sigma^{2}$, 0.01216569, and $\boldsymbol{\beta}$ (see Table \ref{tabla2_ejemplo}) obtained for the OLS after centering the dependent variable (thus, the goodness of fit coincides with the coefficient of determination traditionally applied) will be used.

\begin{algorithm}
 \begin{algorithmic}[1]
     \REQUIRE Calculate $\mathbf{K}$, $\boldsymbol{\Omega}$, $\boldsymbol{\Psi}$, $\boldsymbol{\Theta}$ and $D(n)$ := \{ discretization of the interval [0,1] with $n$ points \}
     \STATE j = 1
     \FOR {$k \in D(n)$}
        \STATE Calculate $MSE \left( \widehat{\boldsymbol{\beta}}(k) \right)$ with expression (\ref{mse_ols}) and save in $mse_{j}$
        \IF {$j > 1$}
            \IF {$mse_{j} > mse_{j-1}$}
                \STATE index = j-1
                \STATE break
            \ENDIF
        \ENDIF
        \STATE j = j + 1
     \ENDFOR
     \STATE $k_{min} = D[index]$
 \end{algorithmic}
 \caption{Obtention of the $k$, $k_{min}$, that minimizes $MSE \left( \widehat{\boldsymbol{\beta}}(k) \right)$}\label{algoritmo}
\end{algorithm}

    Note that it should be verified that $k_{min} < k_{HK}$ since $k_{HK}$ is the maximum threshold established in \cite{HoerlKennard1970b} to be verified that a value of $k$ exists such that $MSE \left( \widehat{\boldsymbol{\beta}}(k) \right) < MSE \left( \widehat{\boldsymbol{\beta}} \right)$. If this situation is not verified, $k_{min} = 0.00083 > 0.0007048761 = k_{HK}$, can be caused by the fact that $\hat{\sigma}^{2}$ is used in the calculation of $k_{HK}$,  while the condition is established for $\sigma^{2}$.

    Furthermore, Table \ref{tabla1_ejemplo} shows the value of $k_{l,min}$ for $l=1,\dots,11$, its MSE and whether it is verified that the MSE is always lower than the one obtained by the OLS. Note that the lowest MSE is obtained for $l=10$, and in this case, the MSE will always be lower than that obtained from the OLS for any value of $k_{10} > 0$. The minimum MSE is obtained for $k_{10,min} = 0.07706729$.
    Note that the second column of this table shows the values of $k_{i}$ proposed by \cite{HoerlKennard1970b}. Thus, the optimal values suggested by Hoerl and Kennard correspond to the one that minimizes the MSE when it is considered that all values of $k_{1},\dots,k_{p}$ are zero except for one of them.

    \begin{table}
        \centering
        \begin{tabular}{cccc}
            \hline
            $l$ & $k_{l,min}$ & $MSE \left( \widehat{\boldsymbol{\beta}}(k_{l,min}) \right)$ &  $\xi_{l} - \sigma^{2}/\lambda_{l} < 0$ \\
            \hline
            1  &  5.675967 $\cdot 10^{7}$ &  2.678111 &   TRUE \\
            2  &  5.130849 $\cdot 10^{3}$ &  2.678111 &  FALSE \\
            3  &  27.21801 &  2.678110 &  FALSE \\
            4  &  7.085955 &  2.678110 &  FALSE \\
            5  &  1.889662 &  2.678110 &  FALSE \\
            6  &  84.11808 &  2.678053 &  FALSE \\
            7  &  3.968410 $\cdot 10^{3}$ &  2.676016 &   TRUE \\
            8  &  7.126586 $\cdot 10^{-2}$ &  2.677647 &  FALSE \\
            9  &  1.754329 $\cdot 10^{-2}$ &  2.670259 &  FALSE \\
            10 &  7.706729 $\cdot 10^{-2}$ &  2.093025 &   TRUE \\
            11 &  7.048761 $\cdot 10^{-4}$ &  2.494481 &  FALSE \\
            \hline
        \end{tabular}
        \caption{$k_{l,min}$ election} \label{tabla1_ejemplo}
    \end{table}

\begin{figure}
          \centering
          \includegraphics[width=10cm]{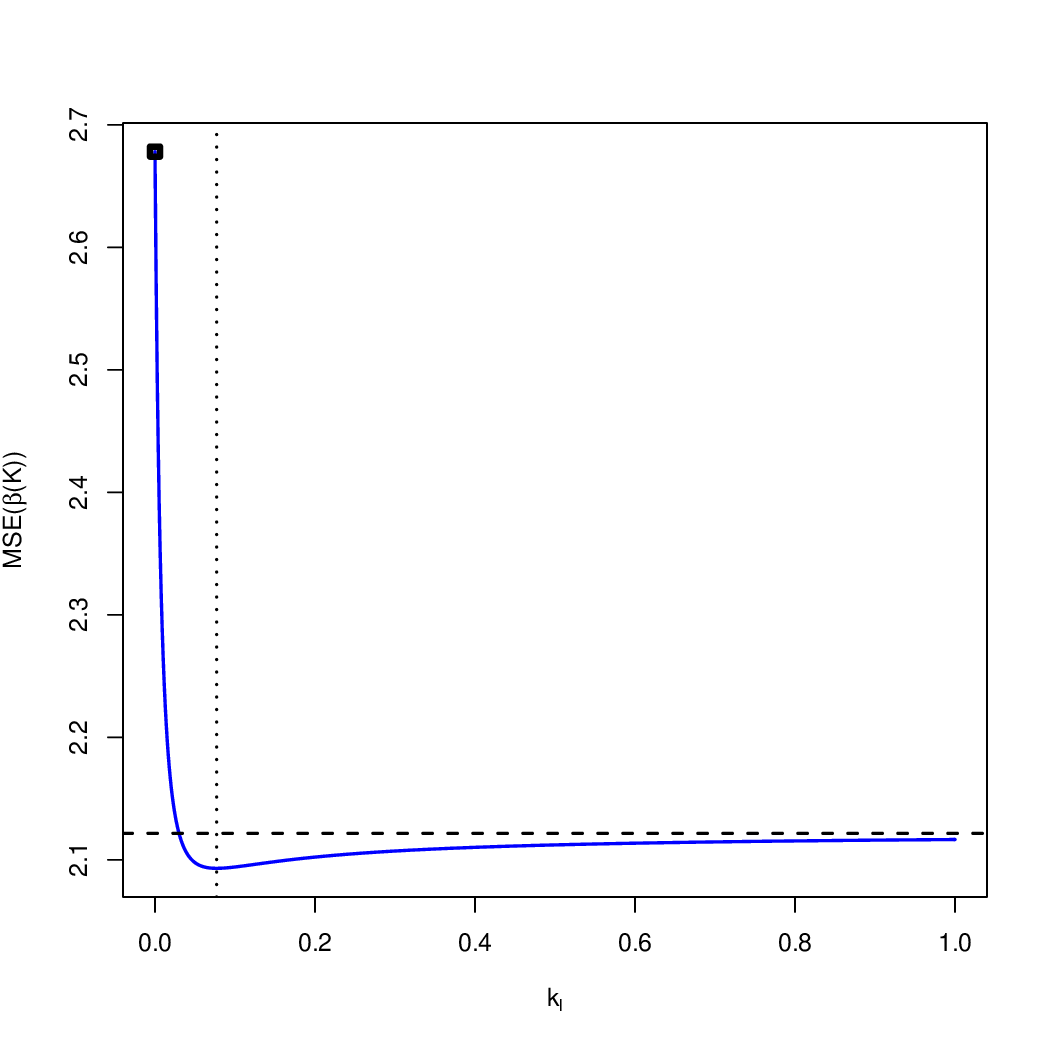}\\
          \caption{Trace for $MSE \left( \widehat{\boldsymbol{\beta}}(k_{10}) \right)$ for $k_{10} \in [0, 1]$. The black point represents $MSE \left( \widehat{\boldsymbol{\beta}} \right)$, the vertical line represents $k_{10,min}$, and horizontal line represents the asymptote $MSE \left( \widehat{\boldsymbol{\beta}} \right) + \xi_{10}^{2} - \widehat{\sigma}^{2}/\lambda_{10}$} \label{fig1}
    \end{figure}

   From the results summarized in Table \ref{tabla2_ejemplo}, it is possible to conclude that the estimation with the lowest MSE is the one obtained when $\mathbf{K} = diag(k_{1},\dots,k_{11})$ for $k_{i} = \frac{\sigma^{2}}{\xi_{i}^{2}}$ with $i=1,\dots,11$, followed by the case where $\mathbf{K} = diag(0,\dots,k_{10,min},\dots,0)$. As was previously commented, in this second case, it is verified that $MSE \left( \widehat{\boldsymbol{\beta}}(k_{10}) \right) < MSE \left( \widehat{\boldsymbol{\beta}} \right)$ for any value of $k_{10}$ (see Figure \ref{fig1}). Furthermore, this estimation is the most similar to the one obtained from the OLS  (with lowest $|| \widehat{\boldsymbol{\beta}} - \widehat{\boldsymbol{\beta}}(\mathbf{K}) ||$).

    \begin{sidewaystable}
        \centering
        \begin{tabular}{ccccccc}
            \hline
            $\mathbf{K}$ & OLS & $k_{HKB} = 0.007316662$ & $k_{HB} = 0.0007048761$ & $k_{min} = 0.00083$ & $k_{i} = \frac{\sigma^{2}}{\xi_{i}^{2}}$ & $k_{10,min} = 0.07706729$ \\
            \hline
            $\widehat{\beta}_{1}$ &  -1.1480402485 &  -0.615975316 &  -1.0558162341 &   -1.0411181661 & -0.7536100103  &  -0.8289615831 \\
            $\widehat{\beta}_{2}$ &  -0.0281064758 &  -0.028590426 &  -0.0281255168 &   -0.0281304843 & -0.0296059930  &  -0.0309439917 \\
            $\widehat{\beta}_{3}$ &  \textbf{-0.0109609943} &  -0.010387826 &  -0.0108660148 &   -0.0108508010 & -0.0095889300  &  -0.0108340116 \\
            $\widehat{\beta}_{4}$ &  \textbf{-0.9948352689} &  -0.899367297 &  -0.9803653295 &   -0.9780152042 & -0.8959178060  &  -0.9926400826 \\
            $\widehat{\beta}_{5}$ &  -0.0546405548 &  -0.057234825 &  -0.0552104328 &   -0.0552980693 & -0.0495166302  &  -0.0545627458 \\
            $\widehat{\beta}_{6}$ &  \textbf{-3.9596038257} &  -1.825723658 &  -3.5638578763 &   -3.5016107255 & -3.6255322448  &  -4.0218644743 \\
            $\widehat{\beta}_{7}$ &   \textbf{0.5449012650} &   0.415759276 &   0.5210035568 &    0.5172413161 &  0.4999095608 &   0.5316978673 \\
            $\widehat{\beta}_{8}$ &   \textbf{0.0278180802} &   0.018243272 &   0.0261355566 &    0.0258683518 &  0.0215278846 &   0.0248643709 \\
            $\widehat{\beta}_{9}$ &   \textbf{0.0480904082} &   0.049696522 &   0.0484754107 &    0.0485336645 &  0.0484407896 &   0.0456378608 \\
            $\widehat{\beta}_{10}$ &  0.0008690746 &   0.001331381 &   0.0009551638 &    0.0009686944 &  0.0007518084 &   0.0008365183  \\
            $\widehat{\beta}_{11}$ &  0.0075720370 &   0.007590831 &   0.0075480354 &    0.0075449443 &  0.0103226880 &   0.0080287843  \\
            \hline
            $|| \widehat{\boldsymbol{\beta}} - \widehat{\boldsymbol{\beta}}(\mathbf{K}) ||$ &  & 4.862431 & 0.1659039 & 0.2222425 & 0.2790656 & 0.1058898 \\
            $MSE$ & 2.678111 & 5.708535 & 2.438379 & 2.433703 & 1.898926 & 2.093025 \\
            $GoF$ & 0.8966053 & 0.8857528 & 0.8962376 & 0.8961127 & 0.8932614 & 0.8959923 \\
            \hline
        \end{tabular}
        \caption{Calculation for the estimation of the generalized ridge and its mean squared error for different possible values of $\mathbf{K}$. Coefficients significantly different from zero are highlighted in bold.} \label{tabla2_ejemplo}
    \end{sidewaystable}

   Unlike in \cite{HoerlKennard1970a}, in this case, no change is observed between the estimated sign of any regressor, which can be because in this case, the data have not been transformed (Hoerl and Kennard considered $\mathbf{X}^{t} \mathbf{X}$ and $\mathbf{X}^{t} \mathbf{Y}$ in correlation form). This fact is observed in the magnitude of the eigenvalues obtained:
    $$\lambda_{1} = 528398.9, \
    \lambda_{2} = 32899.95, \
    \lambda_{3} = 951.4839, \
    \lambda_{4} = 362.2351, \
    \lambda_{5} = 162.681,$$
    $$\lambda_{6} = 98.19544, \
    \lambda_{7} = 5.799103, \
    \lambda_{8} = 1.332221, \
    \lambda_{9} = 0.1563322, \
    \lambda_{10} = 0.01702985, \
    \lambda_{11} = 0.0064903.$$

\begin{figure}
          \centering
          \includegraphics[width=8cm]{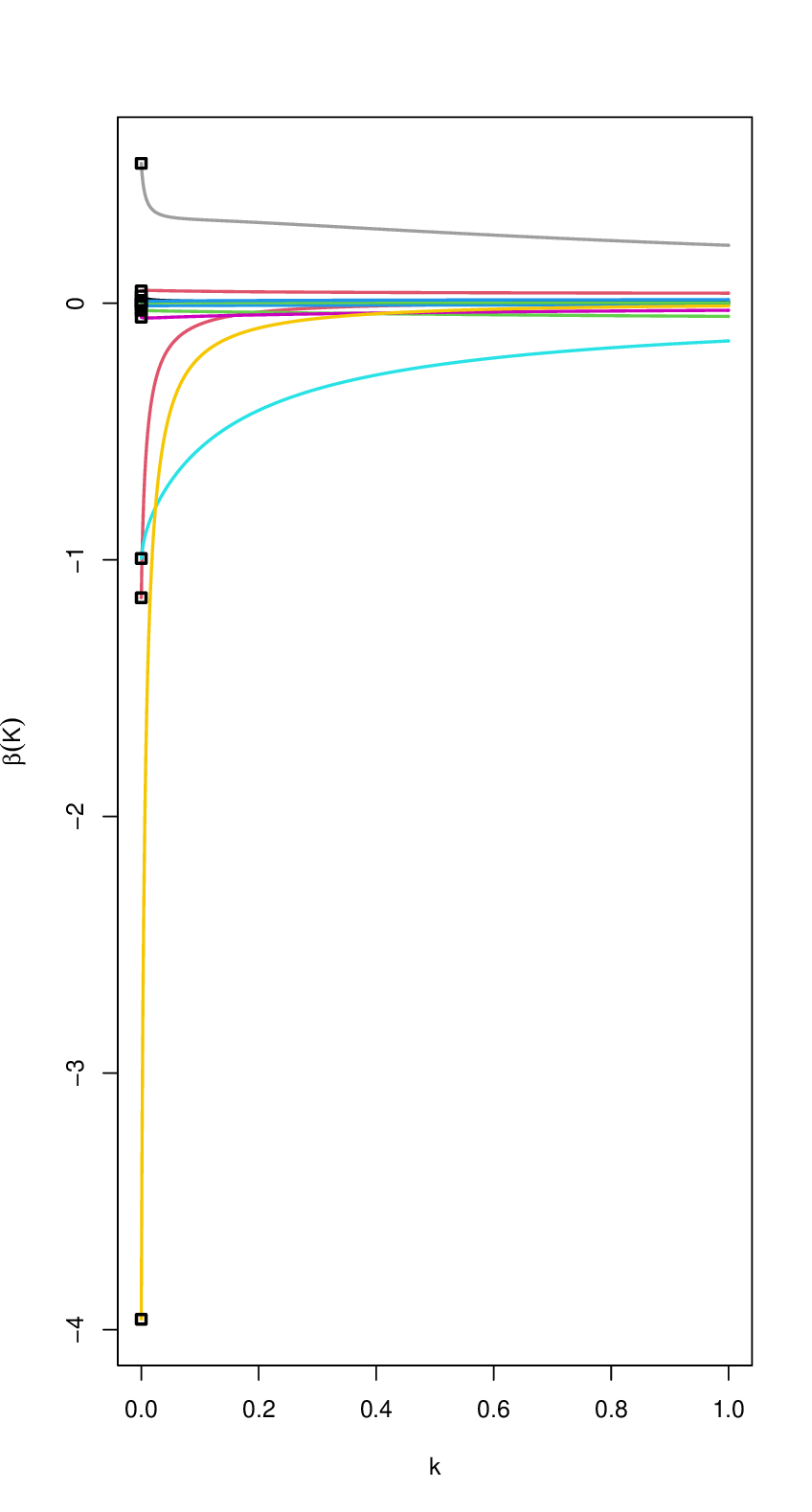}
          \includegraphics[width=8cm]{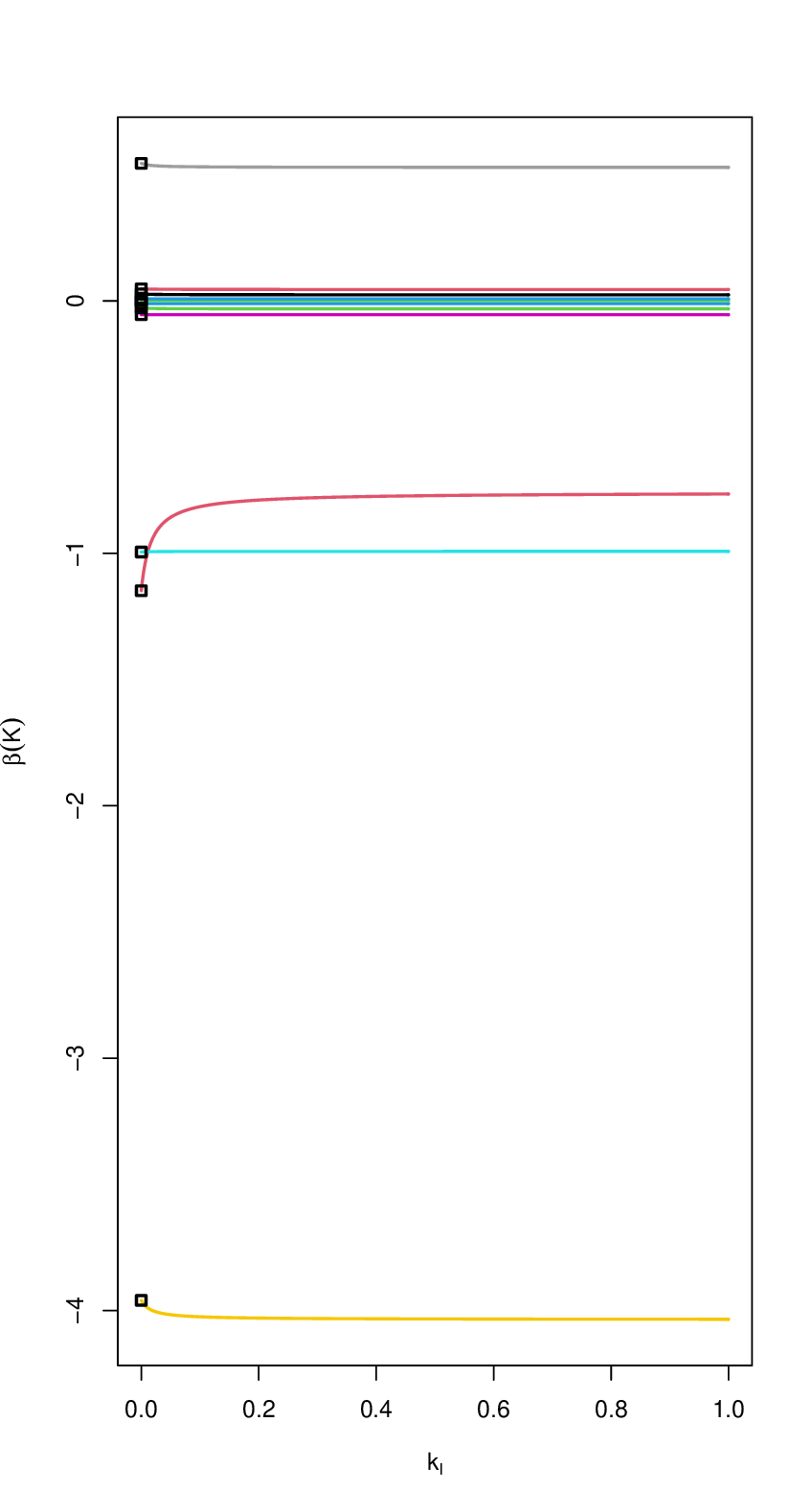}\\
          \caption{Ridge trace for $\widehat{\boldsymbol{\beta}}(k)$ (left) and $\widehat{\boldsymbol{\beta}}(k_{10})$ (right) for $k, k_{10} \in [0, 1]$} \label{fig2}
    \end{figure}

\begin{figure}
          \centering
          \includegraphics[width=10cm]{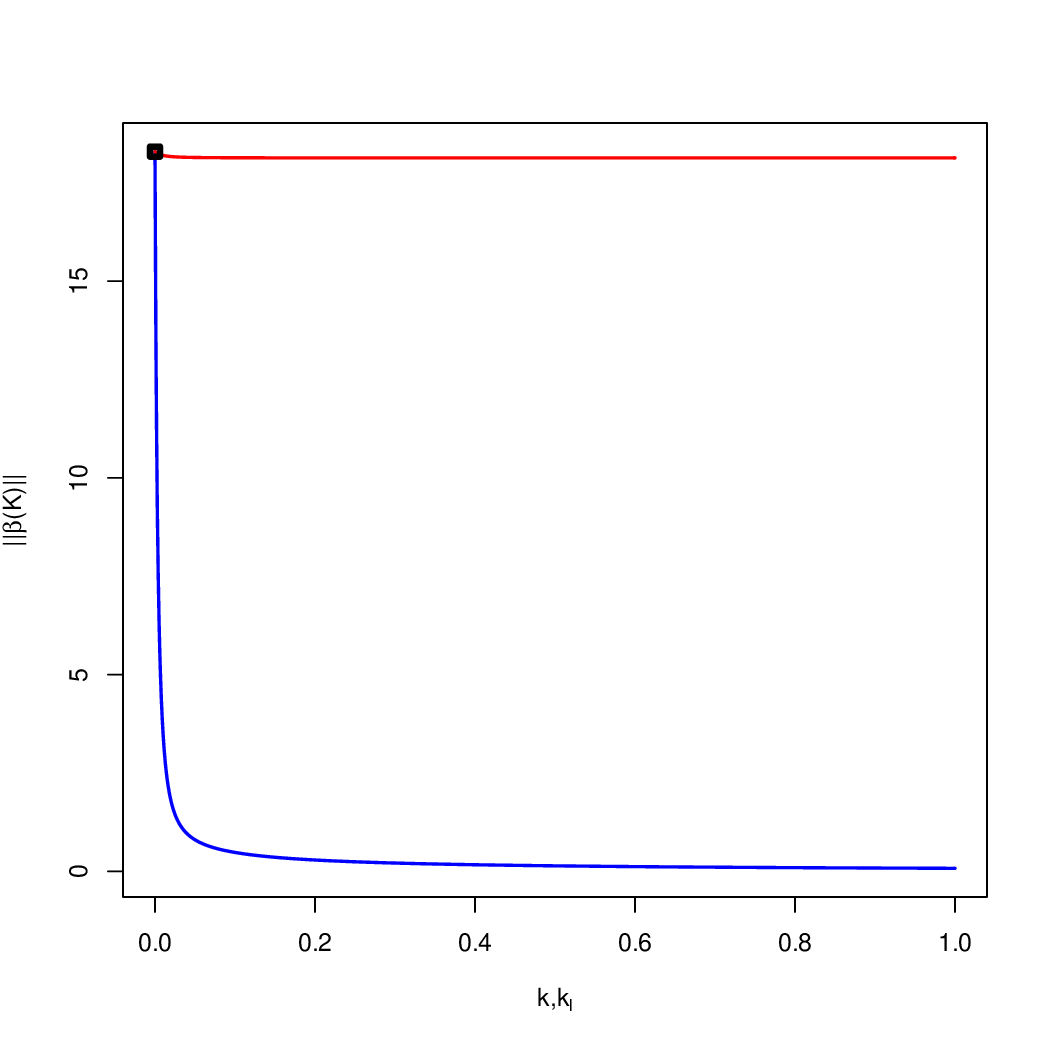}\\
          \caption{Trace for $|| \widehat{\boldsymbol{\beta}}(k) ||$ (blue) and $|| \widehat{\boldsymbol{\beta}}(k_{10}) ||$ (red) for $k, k_{10} \in [0, 1]$. The black point represent $|| \widehat{\boldsymbol{\beta}} ||$} \label{fig3}
    \end{figure}

\begin{figure}
          \centering
          \includegraphics[width=10cm]{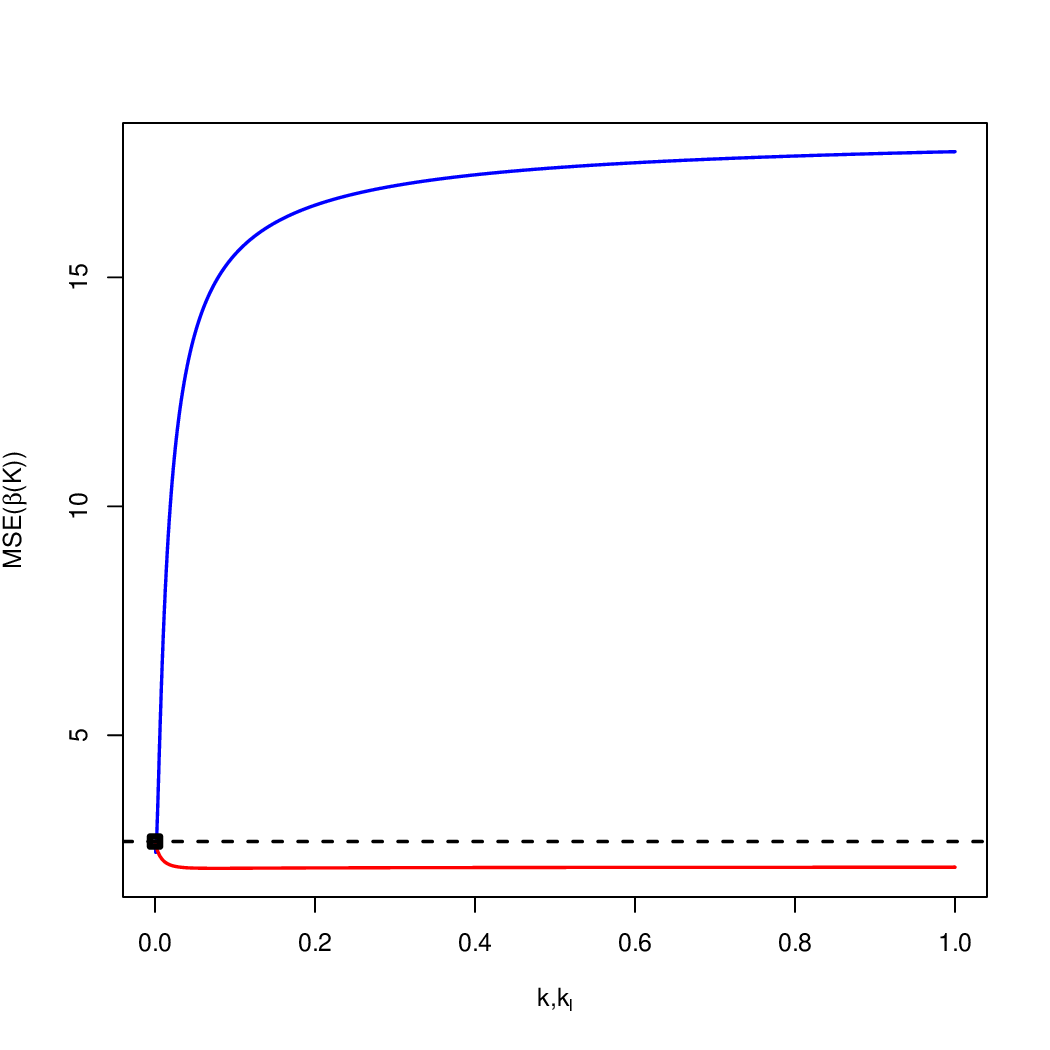}\\
          \caption{Trace for $MSE \left( \widehat{\boldsymbol{\beta}}(k) \right)$ (blue) and $MSE \left( \widehat{\boldsymbol{\beta}}(k_{10}) \right)$ (red) for $k, k_{10} \in [0, 1]$. The black point and horizontal line represent $MSE \left( \widehat{\boldsymbol{\beta}} \right)$} \label{fig4}
    \end{figure}

\begin{figure}
          \centering
          \includegraphics[width=10cm]{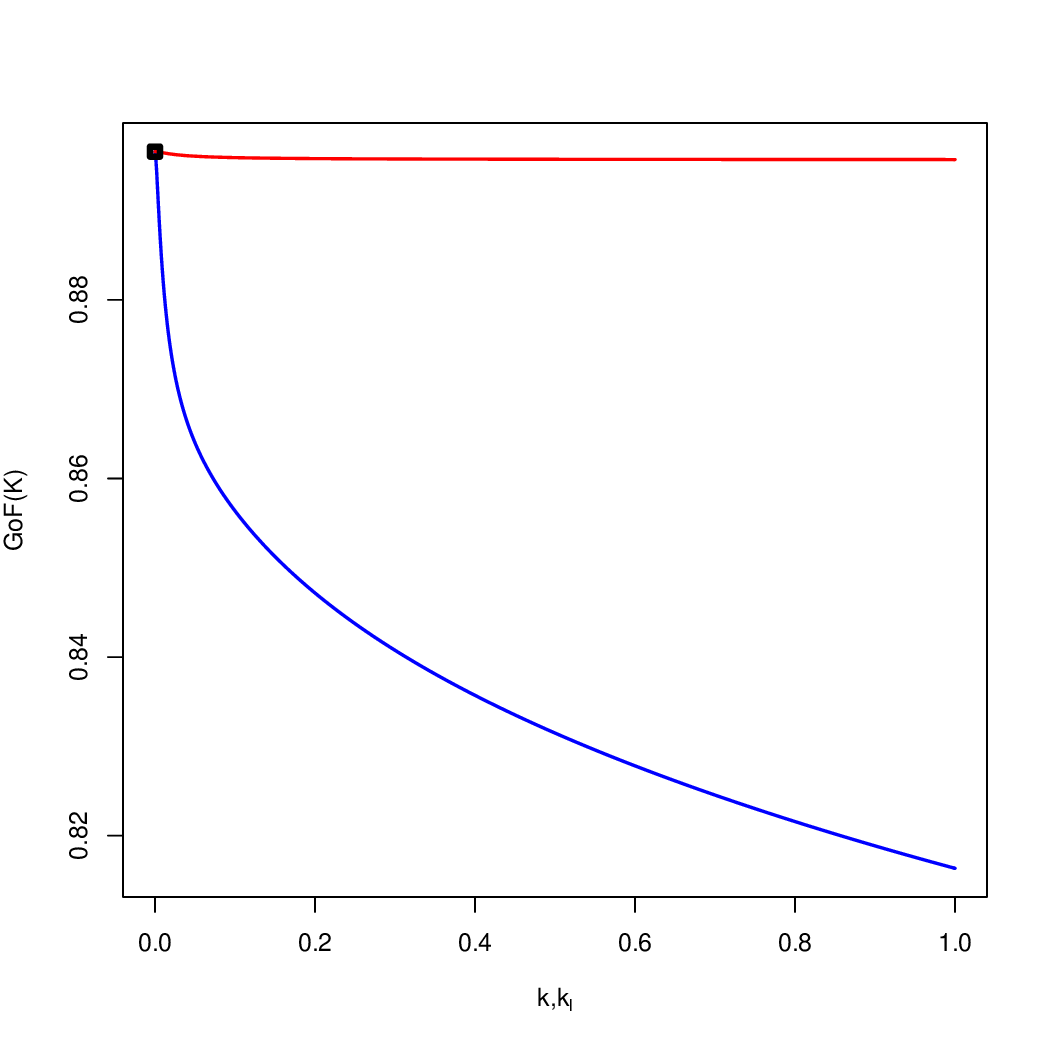}\\
          \caption{Trace for $GoF(k)$ (blue) and $GoF(k_{10})$ (red) for $k, k_{10} \in [0, 1]$. The black point represents $GoF$ for OLS} \label{fig5}
    \end{figure}

   Figure \ref{fig2} shows the trace for the regular and general estimators for $k, k_{10} \in [0,1]$. Note that for the regular case, the estimations converge quickly to zero; while in the generalized case, stability exists around the OLS estimator (see Figure \ref{fig3}). Thus, although the case $\mathbf{K} = diag(0,\dots,k_{10},\dots,0)$ is not useful to select an optimal subset of variables (objective of \cite{GormanToman}), it allows the obtention of an estimation with lower MSE than the one obtained from the OLS and from the regular ridge estimator (see Figure \ref{fig4}). In addition, its goodness of fit is quite superior to the regular case (see Figure \ref{fig5}).

        Finally:
        \begin{itemize}
            \item From Proposition \ref{proposition2}, the generalized ridge estimator $\widehat{\boldsymbol{\beta}}(\mathbf{K})$ with $\mathbf{K}$ given by option c) is preferred over the regular ridge estimator $\widehat{\boldsymbol{\beta}}(k)$ with $k = k_{HB}$ under the root mean squared error matrix criterion since $k_{i} > k$ for all $i=1,\dots,10$ and $k_{11} = k_{HB}$ (see Table \ref{tabla1_ejemplo}).
            \item From Corollary \ref{corolario1}, the generalized ridge estimator $\widehat{\boldsymbol{\beta}}(\mathbf{K})$ with $\mathbf{K}$ given by options c) and d) is preferred over the OLS estimator under the root mean squared error matrix criterion since $k_{i} > 0$ for all $i=1,\dots,11$ (see Table \ref{tabla1_ejemplo}).
            \item From Corollary \ref{corolario2}, the regular ridge estimator $\widehat{\boldsymbol{\beta}}(k)$ with $k = k_{HKB}, k_{HB}, k_{min}$ given by option b) is preferred over the OLS estimator under the root mean squared error matrix criterion since $k_{HKB}, k_{HB}, k_{min} > 0$.
            \item From Corollary \ref{corolario3}, it cannot be stated that the regular ridge estimator $\widehat{\boldsymbol{\beta}}(k)$ with $k = k_{HKB}, k_{HB}, k_{min}$ given by option b) is preferred over the generalized ridge estimator $\widehat{\boldsymbol{\beta}}(\mathbf{K})$ with $\mathbf{K}$ given by option d) under the root mean squared error matrix criterion since $k_{HKB}, k_{HB}, k_{min} \not> k_{10,min}$.
        \end{itemize}

    \subsection{Bootstrap inference and comparison with R packages for regular ridge regression}

        Considering the steps described in section \ref{boots}, Table \ref{tabla3_ejemplo} shows the confidence regions for the coefficient estimates and goodness-of-fit presented in \ref{tabla2_ejemplo}. It is observed that:
        \begin{itemize}
            \item For $k = 0, k_{HK}, k_{min}$ and $\mathbf{K} = diag(0, \dots, k_{10,min}, \dots, 0)$ the same coefficients significantly different from zero are found as in OLS (see coefficients highlighted in bold in Table \ref{tabla2_ejemplo}), it is to say, $\beta_{3}$, $\beta_{4}$, $\beta_{6}$, $\beta_{7}$, $\beta_{8}$ and $\beta_{9}$.
            \item For $k = k_{HKB}$ and $\mathbf{K} = diag(k_{1}, \dots, k_{p})$ with $k_{i} = \frac{\sigma^{2}}{\xi_{i}^{2}}$ for $i=1,\dots,11$, the coefficients significantly different from zero are $\beta_{4}$, $\beta_{6}$, $\beta_{7}$ and $\beta_{9}$.
        \end{itemize}
        It is noteworthy that  \cite{HoerlKennard1970b} proposed to eliminate factors 1, 4, 9 and 10 and that in all the above cases the coefficients $\beta_{4}$ and $\beta_{9}$ are found to be significantly different from zero in all the cases considered.

         In our analysis, it is obtained that the coefficients not significantly different from zero in all the cases are $\beta_{1}$, $\beta_{2}$, $\beta_{5}$, $\beta_{10}$ and $\beta_{11}$. Therefore, citing \cite{HoerlKennard1970b}, \textit{the best subset of size six} would be formed by factors 3, 4, 6, 7, 8 and 9.

    \begin{sidewaystable}
        \centering
        \begin{tabular}{cccc}
            \hline
            $\mathbf{K}$ & OLS & $k_{HKB} = 0.007316662$ & $k_{HB} = 0.0007048761$ \\
            \hline
            $\widehat{\beta}_{1}$ &  (-3.89593670877286,     1.31771775578935) & (-1.4972235906772,      0.80361439048618)    &  (-3.19279865166447,     1.19026203604696) \\
            $\widehat{\beta}_{2}$ &  (-0.0746860210650758,   0.0288701728558158) & (-0.0681295066239293,   0.0226840137910741)  &  (-0.0722700019542601,   0.0272846188871276) \\
            $\widehat{\beta}_{3}$ &  \textbf{(-0.0201978910139842,   -0.00127620689098894)} & (-0.0194915586692353,   0.000678416531221173)  & \textbf{(-0.0200400342689016,   -0.00100072216525185)}  \\
            $\widehat{\beta}_{4}$ &  \textbf{(-1.78077495035861,     -0.166394171769138)}  & \textbf{(-1.57061997928062,     -0.228491939197713)}  & \textbf{(-1.72844833947927,     -0.194212644240152)}  \\
            $\widehat{\beta}_{5}$ &  (-0.187647062123147,    0.0957148698770658)  &  (-0.172852329154543,    0.102080145101977) &  (-0.180611095797693,    0.0983436130632711)  \\
            $\widehat{\beta}_{6}$ &  \textbf{(-8.00272225783783,     -0.948445296629561)}  & \textbf{(-2.71473031693527,     -0.295195799735598)}  & \textbf{(-6.41502094120142,     -0.818074617386231)}   \\
            $\widehat{\beta}_{7}$ &   \textbf{(0.186308069223282,     0.881008395563819)}   &  \textbf{(0.0888668647118551,    0.618211914112481)} &  \textbf{(0.167878220326808,     0.793955175670512)}  \\
            $\widehat{\beta}_{8}$ &   \textbf{(0.000823469538757971,  0.0615556176023861)}  & (-0.000679711901738476, 0.0311983206337429)  & \textbf{(0.000838371938236545,  0.0517612679286025)}   \\
            $\widehat{\beta}_{9}$ &   \textbf{(0.0221546964556036,    0.0791226146581368)}  & \textbf{(0.023721596560448,     0.0741895704635099)}  &  \textbf{(0.0228830017842054,    0.0764930491373058)} \\
            $\widehat{\beta}_{10}$ &  (-0.00220235044946133,  0.0027100927479362) & (-0.000392435367310374, 0.00394377007346599)  &  (-0.00162247498865634,  0.00299335075577762) \\
            $\widehat{\beta}_{11}$ &  (-0.00556691271200455,  0.0222622756193275) & (-0.00462494988856312,  0.0206533525888555)   &  (-0.00513106191290476,  0.0213688925964588) \\
            %\hline
            $GoF$ & (0.857185313518575,     0.97164950986312) & (0.840186693183264,     0.959583929196788)  & (0.856311324646987,     0.97068418568284)  \\
            \hline
        \end{tabular} \\
        \begin{tabular}{cccc}
            \hline
            $\mathbf{K}$ & $k_{min} = 0.00083$ & $k_{i} = \frac{\sigma^{2}}{\xi_{i}^{2}}$ & $k_{10,min} = 0.07706729$ \\
            \hline
            $\widehat{\beta}_{1}$ & (-3.09239223213121,     1.16981841814523)  &  (-3.38520846820602,     1.0512081553846) & (-4.2798251137177,      1.31998326926375) \\
            $\widehat{\beta}_{2}$ & (-0.0720514870609311,   0.0270330434047115)  & (-0.067972207973036,    0.046076825523836)  & (-0.0756048479419764,   0.0272571255591297) \\
            $\widehat{\beta}_{3}$ & \textbf{(-0.0200072077286324,   -0.00097725498919722)}  & (-0.0175438193378674,   0.00464366168337011)  & \textbf{(-0.0200130463050139,   -0.000997629197896326)} \\
            $\widehat{\beta}_{4}$ & \textbf{(-1.7234275126921,      -0.19921955432199)}  &  \textbf{(-1.41926424337071,     -0.11606510002292)} & \textbf{(-1.73791357943147,     -0.164450357806185)} \\
            $\widehat{\beta}_{5}$ & (-0.18001713544389,     0.0983404685195195)  &  (-0.201733516738824,    0.0812856776180331) & (-0.183661133705017,    0.101420519265672) \\
            $\widehat{\beta}_{6}$ & \textbf{(-6.21419416571285,     -0.794509459423308)}  & \textbf{(-6.10934913342415,     -0.526855942293644)}  & \textbf{(-7.61576909348567,     -0.620743075198675)} \\
            $\widehat{\beta}_{7}$ & \textbf{(0.165337227615227,     0.784035140713891)}  & \textbf{(0.158331109934828,     0.740149897755549)}  & \textbf{(0.180033452471868,     0.866413728317379)} \\
            $\widehat{\beta}_{8}$ & \textbf{(0.000810084536847535,  0.0505358206712533)}  & (-0.00499451597678027,  0.0521926397597579)  & \textbf{(0.0000545730315833507,  0.063577654448548)} \\
            $\widehat{\beta}_{9}$ & \textbf{(0.0228902546019822,    0.0764945473119876)}  & \textbf{(0.00324670854793468,   0.0684720229516689)}  & \textbf{(0.022275040865436,     0.0766410723880503)} \\
            $\widehat{\beta}_{10}$ & (-0.0015571902348841,   0.00304229881108438)  & (-0.00526014539785334,  0.00176650694974561)  & (-0.00198860640841325,  0.00286862922654762) \\
            $\widehat{\beta}_{11}$ & (-0.0051988219960957,   0.0212521270327134)  &  (-0.00216000114030527,  0.0273869174467935) & (-0.00551939277398953,  0.022020812221216) \\
            %\hline
            $GoF$ &  (0.856170314838873,     0.970331026132404) & (0.676427685731761,     0.941501956648788) & (0.847262409158761,     0.968610582604451) \\
            \hline
        \end{tabular}
        \caption{Confidence regions for all values in Table \ref{tabla2_ejemplo}. For the coefficient estimates, regions that do not contain zero are highlighted in bold.} \label{tabla3_ejemplo}
    \end{sidewaystable}

        Next, the information shown in Tables \ref{tabla2_ejemplo} and \ref{tabla3_ejemplo} for the regular ridge estimator is compared with the estimation and inference obtained by the R packages  \textit{lmridge} (\cite{lmridge}) and \textit{lrmest} (\cite{lrmest}) de R, which is presented in Table \ref{tabla4_ejemplo}. Note that the package \textit{lrmest}\footnote{
            The command \textit{rid} provides the estimation and inference of the model and the mean squared error
        }:
        \begin{itemize}
            \item provides the same estimations for the coefficients and values of MSE than the one shown in Table \ref{tabla2_ejemplo} for the regular ridge regression.
            \item From Table \ref{tabla5_ejemplo}, exactly the same coefficients significantly different from zero are identified except for the case $k=k_{HKB}$, where it further considers that $\widehat{\beta}_{3}$ and $\widehat{\beta}_{10}$ are significantly different from zero.
        \end{itemize}
        While the package  \textit{lmridge}\footnote{
            The command \textit{lmridge} provides the model estimates including inference and goodness-of-fit, among other values. The command \textit{rstats1} provides, among other values, the mean square error. Finally, the command \textit{kest} provides different estimates for the parameter $k$, including $k_{HKB}$.
        }:
        \begin{itemize}
            \item It provides the same estimates of the coefficients as those given in Table \ref{tabla2_ejemplo} for the regular ridge estimator only when $k=0$, for all other values of $k$ the estimates are not the same but are similar.
            \item The same applies to the goodness-of-fit: it only matches for $k=0$.
            \item The estimation provided for the proposal presented by \cite{HoerlKennardBaldwin} is $k_{HKB} = 0.00689$, which also differs from the value provided in Table \ref{tabla2_ejemplo}.
            \item Finally, Table \ref{tabla5_ejemplo} Identifies exactly the same coefficients significantly different from zero as the bootstrap inference proposed in the present work and that given by the \textit{lrmest} except for the case when $k=k_{HKB}$, where it identifies significantly different from zero the same coefficients as \textit{lrmest} except for $\widehat{\beta}_{8}$.
        \end{itemize}

    \begin{sidewaystable}
        \centering
        \begin{tabular}{|c|cc|cc|cc|cc|}
            \hline
            \multirow{2}{*}{$\mathbf{K}$} & \multicolumn{2}{c|}{$k=0$} & \multicolumn{2}{c|}{$k_{HKB} = 0.007316662$} & \multicolumn{2}{c|}{$k_{HB} = 0.0007048761$} & \multicolumn{2}{c|}{$k_{min} = 0.00083$} \\
                & lrmest & lmridge & lrmest & lmridge & lrmest & lmridge & lrmest & lmridge \\
            \hline
            $\widehat{\beta}_{1}$ & -1.1480 (0.1980) & -1.1480 (0.3007) & -0.6160 (0.3083) & -0.8899 (0.4015)  & -1.0558 (0.2146)  & -1.1018 (0.3125) & -1.0411 (0.2174) & -1.0944 (0.3147) \\
            $\widehat{\beta}_{2}$ & -0.0281 (0.1199)  & -0.0281 (0.1131) & -0.0286 (0.0969) & -0.0264 (0.1486)  & -0.0281 (0.1170)  & -0.0278 (0.1174) & -0.0281 (0.1166) & -0.0277 (0.1182) \\
            $\widehat{\beta}_{3}$ & \textbf{-0.0110} (0.0224)  & \textbf{-0.0110} (0.0201) & \textbf{-0.0104} (0.0291) & \textbf{-0.0105} (0.0307)  & \textbf{-0.0109} (0.0234)  &  \textbf{-0.0109} (0.0210) & \textbf{-0.0109} (0.0236) & \textbf{-0.0109}  (0.0212) \\
            $\widehat{\beta}_{4}$ & \textbf{-0.9948} (0.0015)  & \textbf{-0.9948} (0.0012) & \textbf{-0.8994} (0.0022) & \textbf{-0.9033}  (0.0026)  & \textbf{-0.9804} (0.0016)  & \textbf{-0.9812}  (0.0013) & \textbf{-0.9780} (0.0016) & \textbf{-0.9789}  (0.0014) \\
            $\widehat{\beta}_{5}$ & -0.0546 (0.2555)  & -0.0546 (0.2464) & -0.0572 (0.2328) & -0.0571 (0.2432)  & -0.0552 (0.2505)  & -0.0552 (0.2423) & -0.0553 (0.2497) & -0.0553 (0.2418) \\
            $\widehat{\beta}_{6}$ & \textbf{-3.9596} (0.0071)  & \textbf{-3.9596} (0.0062) & \textbf{-1.8257} (0.0086) & \textbf{-1.8627}  (0.0088)  & \textbf{-3.5639} (0.0073)  & \textbf{-3.5756}  (0.0063) & \textbf{-3.5016} (0.0073) & \textbf{-3.5150}  (0.0063) \\
            $\widehat{\beta}_{7}$ & \textbf{0.5449} (0.0004)  & \textbf{0.5449} (0.0003) & \textbf{0.4158} (0.0009) & \textbf{0.4313}  (0.0011)  & \textbf{0.5210} (0.0004)  & \textbf{0.5239}  (0.0004) & \textbf{0.5172} (0.0004) &  \textbf{0.5206} (0.0004) \\
            $\widehat{\beta}_{8}$ & \textbf{0.0278} (0.0160)  & \textbf{0.278} (0.0142) & \textbf{0.0182} (0.0327) & 0.0210  (0.0526)  & \textbf{0.0261} (0.0179)  & \textbf{0.0266}  (0.0172) & \textbf{0.0259} (0.0183) & \textbf{0.0264} (0.0177) \\
            $\widehat{\beta}_{9}$ & \textbf{0.0481} (0.0007)  & \textbf{0.0481} (0.0006) & \textbf{0.0497} (0.0002) & \textbf{0.0515}  (0.0004)  & \textbf{0.0485} (0.0006)  & \textbf{0.0488}  (0.0005) & \textbf{0.0485} (0.0006) & \textbf{0.0489}  (0.0005) \\
            $\widehat{\beta}_{10}$ & 0.0009 (0.2161)  & 0.0009 (0.2074) & \textbf{0.0013} (0.0449) & \textbf{0.0013}  (0.0480)  & 0.0010 (0.1678)  & 0.0010 (0.1604) & 0.0010 (0.1610) & 0.0010 (0.1539) \\
            $\widehat{\beta}_{11}$ & 0.0076 (0.2594)  & 0.0076 (0.2503) & 0.0076 (0.2530) & 0.0072  (0.2883)  & 0.0075 (0.2601)  & 0.0075 (0.2556) & 0.0075 (0.2602) & 0.0075 (0.2565) \\
            \hline
            $MSE$ & 2.6781  & 1.8505 & 5.7085 & 4.9392  & 2.4384  & 1.6801 & 2.4337 & 1.6839 \\
            $GoF$ &   & 0.896600 &  & 0.864300  &   & 0.889500 &  & 0.88850 \\
            \hline
        \end{tabular}
        \caption{Estimation and inference obtained from the \textit{lmridge} and \textit{lrmest} packages of R. Coefficients significantly different from zero are highlighted in bold (p-value in brackets).} \label{tabla4_ejemplo}
    \end{sidewaystable}

    \begin{sidewaystable}
        \centering
        \begin{tabular}{|c|ccc|ccc|ccc|ccc|}
            \hline
            \multirow{2}{*}{$\mathbf{K}$} & \multicolumn{3}{c|}{$k=0$} & \multicolumn{3}{c|}{$k_{HKB} = 0.007316662$} & \multicolumn{3}{c|}{$k_{HB} = 0.0007048761$} & \multicolumn{3}{c|}{$k_{min} = 0.00083$} \\
                & GRR & lrmest & lmridge & GRR & lrmest & lmridge & GRR & lrmest & lmridge & GRR & lrmest & lmridge \\
            \hline
            $\widehat{\beta}_{1}$ &  &  &  &  &  &  &  &  &  &  &  &  \\
            $\widehat{\beta}_{2}$ &  &  &  &  &  &  &  &  &  &  &  &  \\
            $\widehat{\beta}_{3}$ & $\checkmark$ & $\checkmark$ & $\checkmark$ &  & $\checkmark$ & $\checkmark$ & $\checkmark$ & $\checkmark$ & $\checkmark$ & $\checkmark$ & $\checkmark$ & $\checkmark$ \\
            $\widehat{\beta}_{4}$ & $\checkmark$ & $\checkmark$ & $\checkmark$ & $\checkmark$ & $\checkmark$ & $\checkmark$ & $\checkmark$ & $\checkmark$ & $\checkmark$ & $\checkmark$ & $\checkmark$ &$\checkmark$  \\
            $\widehat{\beta}_{5}$ &  &  &  &  &  &  &  &  &  &  &  &  \\
            $\widehat{\beta}_{6}$ & $\checkmark$ & $\checkmark$ & $\checkmark$ & $\checkmark$ & $\checkmark$ & $\checkmark$ & $\checkmark$ & $\checkmark$ & $\checkmark$ & $\checkmark$ & $\checkmark$ & $\checkmark$ \\
            $\widehat{\beta}_{7}$ & $\checkmark$ & $\checkmark$ & $\checkmark$ & $\checkmark$ & $\checkmark$ & $\checkmark$ & $\checkmark$ & $\checkmark$ & $\checkmark$ & $\checkmark$ & $\checkmark$ & $\checkmark$ \\
            $\widehat{\beta}_{8}$ & $\checkmark$ & $\checkmark$ & $\checkmark$ &  & $\checkmark$ &  & $\checkmark$ & $\checkmark$ & $\checkmark$ & $\checkmark$ & $\checkmark$ & $\checkmark$ \\
            $\widehat{\beta}_{9}$ & $\checkmark$ & $\checkmark$ & $\checkmark$ & $\checkmark$ & $\checkmark$ & $\checkmark$ & $\checkmark$ & $\checkmark$ & $\checkmark$ & $\checkmark$ & $\checkmark$ & $\checkmark$ \\
            $\widehat{\beta}_{10}$ &  &  &  &  & $\checkmark$ & $\checkmark$ &  &  &  &  &  &  \\
            $\widehat{\beta}_{11}$ &  &  &  &  &  &  &  &  &  &  &  &  \\
            \hline
        \end{tabular}
        \caption{Comparison of the individual significance of each coefficient in the different methodologies considered for regular ridge regression.} \label{tabla5_ejemplo}
    \end{sidewaystable}

        It should be noted that the R packages \textit{genridge} (\cite{genridge}) and \textit{ridge} (\cite{ridge}) have also been used, obtaining from the former values very different from those presented in this paper and from the latter exactly the same values as those given by the \textit{lmridge} package. Other R packages that provide estimates for regular ridge regression are listed in \cite{lmridge_paper}. However, in order not to extend the present work, we have considered what we believe to be the most representative.

\section{Conclusions}
    \label{conclusion}

     \cite{HoerlKennard1970a,HoerlKennard1970b} presented the ridge estimation for the particular case when $\mathbf{K} = k  \mathbf{I}$ (known as the regular ridge estimator), although they parted from a general case in which $\mathbf{K}$ is a diagonal matrix whose elements can be all different between them. This paper develops this alternative version of the general case that was not previously analyzed to the best of our knowledge. We pay special attention to the case in which all the elements of the diagonal of matrix $\mathbf{K}$ are equal to zero except for one, $k_{l}$, with $l=1,\dots,p$.

    As a relevant contribution, this paper presents the expression of this general estimator, which is different from the one presented by Hoerl and Kennard. This paper also analyzed the estimator's main characteristics (unbiased, matrix of variances and covariances and the augmented model), its norm, its mean squared error and its goodness of fit. The expressions obtained for the norm, mean squared error and goodness of fit verify its property of being continuous (i.e., coincides with the expressions of the OLS when $\mathbf{K}$ is a null matrix). As would be desirable, the norm and the measure of goodness of fit decrease as a function of $\mathbf{K}$.

    In relation to the particular case when $\mathbf{K} = diag(0,\dots,k_{l},\dots,0)$, the following is observed:
    \begin{itemize}
        \item All the elements of matrix $\mathbf{X}^{t} \mathbf{X}$ are affected in the calculation of this estimator instead of in the case of the regular estimator regular, where only the elements of the main diagonal are affected. It could be interesting to analyze whether this generalization improves the calculation of the inverse matrix $\mathbf{X}^{t} \mathbf{X}$ in comparison with the regular case. In addition, and contrary to the regular case, the estimations do not converge towards zero but are around the OLS estimation.
        \item The norm of the estimator decreases and converges around the norm of the OLS estimator (again, it is not converging towards zero as in the regular case). This fact indicates that a range of values for $k_{l}$ that stabilize the calculated estimation can exist.
        \item Contrary to the regular case, it is possible to calculate the value of $k_{l}$ that minimizes the MSE and that leads to a MSE lower than that obtained from the OLS. From the two scenarios obtained when analyzing its asymptotic behavior, in one of them, the MSE is always lower than the one obtained from OLS regardless of the value of $k_{l}$.
        \item A new original alternative for measuring the goodness of fit not only in this generalization but also in the regular case is proposed. The closed expression was obtained and analyzed being decreasing as a function of $k_{l}$. When the dependent variable has zero mean, this alternative version coincides with the coefficient of determination traditionally applied. For standardized data and for the regular case, it coincides with the proposal presented by \cite{Rodriguez2019R2}.
    \end{itemize}

    In conclusion, this particular case provides higher stability for the calculated expressions. This fact makes it preferable to the other options considered, as shown in the example. In addition, the proposed bootstrap inference identifies those coefficients significantly different from zero.

    Finally, as a future research line, it could be interesting to analyze the usability of this particular case to mitigate the degree of near multicollinearity existing in the multiple linear regression model. In addition, due to the differences detected in the illustrative example regarding the results provided by the different \cite{RCoreTeam} packages considered nd to the fact that none (to the best of our knowledge) has the option of generalized ridge regression, we consider it appropriate to approach the creation of a package that integrates the code provided in Github (\url{https://github.com/rnoremlas/GRR/tree/main/01_Biased_estimation}).

%\section*{Acknowledgments}

%We would like to thank the reviewers for their valuable comments and suggestions.
%This work was supported by project PP2019-EI-02 of the University of Granada (Spain) and project A-SEJ-496-UGR20 I+D+i (FEDER Andaluc\'ia, 2014-2020).

\appendix

\section{Goodness of Fit and Data Transformation}
    \label{appendixGoF}

    Given the expression:
    $$GoF_{\mathbf{Y}} = \frac{\sum \limits_{i=1}^{n} \widehat{Y}_{i}^{2}}{\sum \limits_{i=1}^{n} Y_{i}^{2}},$$
    and by considering the transformation $y_{i} = \frac{Y_{i} - a}{b}$ for $i=1,\dots,n$ with $a, b \in \mathbb{R}-\{0\}$, it is obtained that:
    $$\sum \limits_{i=1}^{n} \widehat{y}_{i}^{2} = \frac{1}{b^{2}} \sum \limits_{i=1}^{n} (\widehat{Y}_{i} - a)^{2}, \quad
    \sum \limits_{i=1}^{n} y_{i}^{2} = \frac{1}{b^{2}} \sum \limits_{i=1}^{n} (Y_{i} - a)^{2},$$
    and, consequently:
    $$GoF_{\mathbf{y}} = \frac{\sum \limits_{i=1}^{n} (\widehat{Y}_{i} - a)^{2}}{\sum \limits_{i=1}^{n} (Y_{i} - a)^{2}} \not= GoF_{\mathbf{Y}}.$$
    It is concluded that the GoF is affected by origin changes but not by scale changes.

\bibliographystyle{chicago}
\bibliography{bib}

\begin{thebibliography}{}

\bibitem[\protect\citeauthoryear{Balakrishnant}{Balakrishnant}{1963}]{balakrishnant1963operator}
Balakrishnant, A. (1963).
\newblock An operator theoretic formulation of a class of control problems and
  a steepest descent method of solution.
\newblock {\em Journal of the Society for Industrial and Applied Mathematics,
  Series A: Control\/}~{\em 1\/}(2), 109--127.

\bibitem[\protect\citeauthoryear{Casella}{Casella}{1980}]{casella1980minimax}
Casella, G. (1980).
\newblock Minimax ridge regression estimation.
\newblock {\em The Annals of Statistics\/}, 1036--1056.

\bibitem[\protect\citeauthoryear{Cule, Moritz, and Frankowski}{Cule
  et~al.}{2022}]{ridge}
Cule, E., S.~Moritz, and D.~Frankowski (2022).
\newblock {\em ridge: Ridge Regression with Automatic Selection of the Penalty
  Parameter}.
\newblock R package version 3.3.

\bibitem[\protect\citeauthoryear{Dissanayake and Wijekoon}{Dissanayake and
  Wijekoon}{2016}]{lrmest}
Dissanayake, A. and P.~Wijekoon (2016).
\newblock {\em lrmest: Different Types of Estimators to Deal with
  Multicollinearity}.
\newblock R package version 3.0.

\bibitem[\protect\citeauthoryear{Donoho and Johnstone}{Donoho and
  Johnstone}{1995}]{donoho1995adapting}
Donoho, D.~L. and I.~M. Johnstone (1995).
\newblock Adapting to unknown smoothness via wavelet shrinkage.
\newblock {\em Journal of the american statistical association\/}~{\em
  90\/}(432), 1200--1224.

\bibitem[\protect\citeauthoryear{Efron}{Efron}{1986}]{EfronTibshirani1986}
Efron, B., T.~R. (1986).
\newblock Bootstrap methods for standard errors, confidence intervals, and
  other measures of statistical accuracy.
\newblock {\em Statistical Science\/}~{\em 1\/}(1), 54--75.

\bibitem[\protect\citeauthoryear{Farebrother}{Farebrother}{1976}]{Farebrother1976}
Farebrother, R. (1976).
\newblock Further results on the mean square error of ridge regression.
\newblock {\em Journal of the Royal Statistical Society. Series B
  (Methodological)\/}, 248--250.

\bibitem[\protect\citeauthoryear{Fletcher}{Fletcher}{2013}]{fletcher2013practical}
Fletcher, R. (2013).
\newblock {\em Practical methods of optimization}.
\newblock John Wiley \& Sons.

\bibitem[\protect\citeauthoryear{Frank and Friedman}{Frank and
  Friedman}{1993}]{frank1993statistical}
Frank, L.~E. and J.~H. Friedman (1993).
\newblock A statistical view of some chemometrics regression tools.
\newblock {\em Technometrics\/}~{\em 35\/}(2), 109--135.

\bibitem[\protect\citeauthoryear{Friendly}{Friendly}{2023}]{genridge}
Friendly, M. (2023).
\newblock {\em genridge: Generalized Ridge Trace Plots for Ridge Regression}.
\newblock R package version 0.7.0.

\bibitem[\protect\citeauthoryear{Fu}{Fu}{1998}]{fu1998penalized}
Fu, W.~J. (1998).
\newblock Penalized regressions: the bridge versus the lasso.
\newblock {\em Journal of computational and graphical statistics\/}~{\em
  7\/}(3), 397--416.

\bibitem[\protect\citeauthoryear{Gorman and Toman}{Gorman and
  Toman}{1970}]{GormanToman}
Gorman, J. and R.~Toman (1970).
\newblock Selection of variables for fitting equations to data.
\newblock {\em Technometrics\/}~{\em 8}, 27--51.

\bibitem[\protect\citeauthoryear{Halawa and El~Bassiouni}{Halawa and
  El~Bassiouni}{2000}]{halawa2000tests}
Halawa, A. and M.~El~Bassiouni (2000).
\newblock Tests of regression coefficients under ridge regression models.
\newblock {\em Journal of Statistical Computation and Simulation\/}~{\em
  65\/}(1-4), 341--356.

\bibitem[\protect\citeauthoryear{Harville}{Harville}{1998}]{harville1998matrix}
Harville, D.~A. (1998).
\newblock Matrix algebra from a statistician's perspective.

\bibitem[\protect\citeauthoryear{Hastie}{Hastie}{2020}]{hastie2020ridge}
Hastie, T. (2020).
\newblock Ridge regularization: An essential concept in data science.
\newblock {\em Technometrics\/}~{\em 62\/}(4), 426--433.

\bibitem[\protect\citeauthoryear{Hoerl}{Hoerl}{1962}]{Hoerl1962}
Hoerl, A. (1962).
\newblock Application of ridge analysis to regression problems.
\newblock {\em Chemical Engineering Progress\/}~{\em 58}, 54--59.

\bibitem[\protect\citeauthoryear{Hoerl}{Hoerl}{1964}]{Hoerl1964}
Hoerl, A. (1964).
\newblock Ridge analysis.
\newblock {\em Chemical Engineering Progress Symposium\/}~{\em Series 60},
  67--77.

\bibitem[\protect\citeauthoryear{Hoerl and Kennard}{Hoerl and
  Kennard}{1968}]{HoerlKennard1968}
Hoerl, A. and R.~Kennard (1968).
\newblock On regression analysis and biased estimation.
\newblock {\em Technometrics\/}~{\em 10\/}(Abstract), 422--423.

\bibitem[\protect\citeauthoryear{Hoerl, Kannard, and Baldwin}{Hoerl
  et~al.}{1975}]{HoerlKennardBaldwin}
Hoerl, A.~E., R.~W. Kannard, and K.~F. Baldwin (1975).
\newblock Ridge regression: some simulations.
\newblock {\em Communications in Statistics-Theory and Methods\/}~{\em 4\/}(2),
  105--123.

\bibitem[\protect\citeauthoryear{Hoerl and Kennard}{Hoerl and
  Kennard}{1970a}]{HoerlKennard1970b}
Hoerl, A.~E. and R.~W. Kennard (1970a).
\newblock Ridge regression: applications to nonorthogonal problems.
\newblock {\em Technometrics\/}~{\em 12\/}(1), 69--82.

\bibitem[\protect\citeauthoryear{Hoerl and Kennard}{Hoerl and
  Kennard}{1970b}]{HoerlKennard1970a}
Hoerl, A.~E. and R.~W. Kennard (1970b).
\newblock Ridge regression: Biased estimation for nonorthogonal problems.
\newblock {\em Technometrics\/}~{\em 12\/}(1), 55--67.

\bibitem[\protect\citeauthoryear{Hoerl}{Hoerl}{2020}]{hoerl2020ridge}
Hoerl, R.~W. (2020).
\newblock Ridge regression: a historical context.
\newblock {\em Technometrics\/}~{\em 62\/}(4), 420--425.

\bibitem[\protect\citeauthoryear{Imdad, Aslam, and Altaf}{Imdad
  et~al.}{2018}]{lmridge_paper}
Imdad, M., M.~Aslam, and S.~Altaf (2018).
\newblock lmridge: A comprehensive r package for ridge regression.
\newblock {\em The R Journal\/}~{\em 10\/}(2), 326--346.

\bibitem[\protect\citeauthoryear{Imdad and Aslam}{Imdad and
  Aslam}{2023}]{lmridge}
Imdad, M.~U. and M.~Aslam (2023).
\newblock {\em {lmridge}: Linear Ridge Regression with Ridge Penalty and Ridge
  Statistics}.
\newblock R package version 1.2.2.

\bibitem[\protect\citeauthoryear{Jeffreys}{Jeffreys}{1998}]{jeffreys1998theory}
Jeffreys, H. (1998).
\newblock {\em The theory of probability}.
\newblock OUP Oxford.

\bibitem[\protect\citeauthoryear{Klinger}{Klinger}{1998}]{klinger1998hochdimensionale}
Klinger, A. (1998).
\newblock {\em Hochdimensionale generalisierte lineare Modelle}.
\newblock Shaker.

\bibitem[\protect\citeauthoryear{Ljndley and Smith}{Ljndley and
  Smith}{1972}]{ljndley1972bayes}
Ljndley, D. and A.~Smith (1972).
\newblock Bayes estimators for the linear model (with discussion).
\newblock {\em JR Statist. Soc\/}, 1--41.

\bibitem[\protect\citeauthoryear{Marquardt}{Marquardt}{1970}]{Marquardt1970}
Marquardt, D.~W. (1970).
\newblock Generalized inverses, ridge regression, biased linear estimation, and
  nonlinear estimation.
\newblock {\em Technometrics\/}~{\em 12\/}(3), 591--612.

\bibitem[\protect\citeauthoryear{Obenchain}{Obenchain}{1975}]{Obenchain1975}
Obenchain, R. (1975).
\newblock Ridge analysis following a preliminary test of the shrunken
  hypothesis.
\newblock {\em Technometrics\/}~{\em 17\/}(4), 431--441.

\bibitem[\protect\citeauthoryear{Obenchain}{Obenchain}{1977}]{obenchain1977classical}
Obenchain, R. (1977).
\newblock Classical f-tests and confidence regions for ridge regression.
\newblock {\em Technometrics\/}~{\em 19\/}(4), 429--439.

\bibitem[\protect\citeauthoryear{Piegorsch and Casella}{Piegorsch and
  Casella}{1989}]{piegorsch1989early}
Piegorsch, W.~W. and G.~Casella (1989).
\newblock The early use of matrix diagonal increments in statistical problems.
\newblock {\em SIAM review\/}~{\em 31\/}(3), 428--434.

\bibitem[\protect\citeauthoryear{{R Core Team}}{{R Core
  Team}}{2022}]{RCoreTeam}
{R Core Team} (2022).
\newblock {\em R: A Language and Environment for Statistical Computing}.
\newblock Vienna, Austria: R Foundation for Statistical Computing.

\bibitem[\protect\citeauthoryear{Raiffa and Schlaifer}{Raiffa and
  Schlaifer}{1961}]{raiffa1961applied}
Raiffa, H. and R.~Schlaifer (1961).
\newblock Applied statistical decision theory.
\newblock Technical report.

\bibitem[\protect\citeauthoryear{Rodr\'iguez, Salmer\'on, and
  Garc\'ia}{Rodr\'iguez et~al.}{2019}]{Rodriguez2019R2}
Rodr\'iguez, A., R.~Salmer\'on, and C.~Garc\'ia (2019).
\newblock The coefficient of determination in the ridge regression.
\newblock {\em Communications in Statistics - Simulation and Computation\/},
  https://doi.org/10.1080/03610918.2019.1649421.

\bibitem[\protect\citeauthoryear{Rodr\'iguez, Salmer\'on, and
  Garc\'ia}{Rodr\'iguez et~al.}{2021}]{Rodriguezetal2021}
Rodr\'iguez, A., R.~Salmer\'on, and C.~Garc\'ia (2021).
\newblock Obtaining a threshold for the stewart index and its extension to
  ridge regression.
\newblock {\em Computational Statistics\/}~{\em 36}, 1011--1029.

\bibitem[\protect\citeauthoryear{Rolph}{Rolph}{1976}]{rolph1976choosing}
Rolph, J.~E. (1976).
\newblock Choosing shrinkage estimators for regression problems.
\newblock {\em Communications in Statistics-Theory and Methods\/}~{\em 5\/}(9),
  789--802.

\bibitem[\protect\citeauthoryear{Salmer\'on, Garc\'ia, and Garc\'ia}{Salmer\'on
  et~al.}{2024}]{Salmeron2024}
Salmer\'on, R., C.~Garc\'ia, and J.~Garc\'ia (2024).
\newblock The raise regression: Justification, properties and application.
\newblock {\em International Statistical Review\/}~{\em Accepted}.

\bibitem[\protect\citeauthoryear{Salmer{\'o}n, Garc{\'\i}a, and
  Garc{\'\i}a~P{\'e}rez}{Salmer{\'o}n et~al.}{2020}]{Salmeron2020centered}
Salmer{\'o}n, R., C.~G. Garc{\'\i}a, and J.~Garc{\'\i}a~P{\'e}rez (2020).
\newblock Detection of near-multicollinearity through centered and noncentered
  regression.
\newblock {\em Mathematics\/}~{\em 8}, 931.

\bibitem[\protect\citeauthoryear{Stein}{Stein}{1960}]{stein1960multiple}
Stein, C. (1960).
\newblock Multiple regression, contributions to probability and statistics.
\newblock {\em essays in honor of Harold Hotelling\/}~{\em 103}.

\bibitem[\protect\citeauthoryear{Strawderman}{Strawderman}{1978}]{strawderman1978minimax}
Strawderman, W.~E. (1978).
\newblock Minimax adaptive generalized ridge regression estimators.
\newblock {\em Journal of the American Statistical Association\/}~{\em
  73\/}(363), 623--627.

\bibitem[\protect\citeauthoryear{Theobald}{Theobald}{1974}]{Theobald1974}
Theobald, C.~M. (1974).
\newblock Generalizations of mean square error applied to ridge regression.
\newblock {\em Journal of the Royal Statistical Society: Series B
  (Methodological)\/}~{\em 36\/}(1), 103--106.

\bibitem[\protect\citeauthoryear{Tibshirani}{Tibshirani}{1996}]{tibshirani1996regression}
Tibshirani, R. (1996).
\newblock Regression shrinkage and selection via the lasso.
\newblock {\em Journal of the Royal Statistical Society: Series B
  (Methodological)\/}~{\em 58\/}(1), 267--288.

\bibitem[\protect\citeauthoryear{Trenklar}{Trenklar}{1980}]{Trenklar1980}
Trenklar, G. (1980).
\newblock Generalized mean squared error comparisons of biased regression
  estimators.
\newblock {\em Communications in Statistics-Theory and Methods\/}~{\em
  9\/}(12), 1247--1259.

\bibitem[\protect\citeauthoryear{Zhang and Politis}{Zhang and
  Politis}{2022}]{zhang2022ridge}
Zhang, Y. and D.~N. Politis (2022).
\newblock Ridge regression revisited: Debiasing, thresholding and bootstrap.
\newblock {\em The Annals of Statistics\/}~{\em 50\/}(3), 1401--1422.

\bibitem[\protect\citeauthoryear{Zou and Hastie}{Zou and
  Hastie}{2005}]{zou2005regularization}
Zou, H. and T.~Hastie (2005).
\newblock Regularization and variable selection via the elastic net.
\newblock {\em Journal of the royal statistical society: series B (statistical
  methodology)\/}~{\em 67\/}(2), 301--320.

\end{thebibliography}

\end{document}